\documentclass[apj]{emulateapj}

\usepackage{graphicx}
\usepackage{apjfonts}

\def\arcsec{$^{\prime\prime}$}
\def\ergssec   {~ergs~s$^{-1}$}
\def\kms   {~km~s$^{-1}$}
\def\deg{$^{\circ}$}
\def\msun{{M$_{\odot}$}}

\bibpunct{(}{)}{,}{a}{}{}

\shorttitle{Reflections of AGN Outbursts}
\shortauthors{Forman et al.}

\submitted{Submitted to ApJ -- Dec 21, 2003}

\begin{document}

\title{Reflections of AGN Outbursts in the Gaseous Atmosphere of M87}
\author{W. Forman\altaffilmark{1}, 
P. Nulsen\altaffilmark{1,2},
S. Heinz\altaffilmark{3},
F. Owen\altaffilmark{4},
J. Eilek\altaffilmark{5},
A. Vikhlinin\altaffilmark{1,6},\\
M. Markevitch\altaffilmark{1},
R. Kraft\altaffilmark{1},
E. Churazov\altaffilmark{7,6},
C. Jones\altaffilmark{1}
}

\altaffiltext{1}{Smithsonian Astrophysical Observatory,
Harvard-Smithsonian Center for Astrophysics, 60 Garden St., Cambridge,
MA 02138; wrf@cfa.harvard.edu}

\altaffiltext{2}{On leave from the University of Wollongong}

\altaffiltext{3}{MIT, Cambridge, MA}

\altaffiltext{4}{National Radio Astronomy Observatory, Socorro, NM
87801}

\altaffiltext{5}{New Mexico Tech., Socorro, NM 87801}

\altaffiltext{6}{Space Research Institute (IKI), Profsoyuznaya 84/32,
Moscow 117810, Russia}

\altaffiltext{7}{MPI f\"{u}r Astrophysik, Karl-Schwarzschild-Strasse
1, 85740
Garching, Germany}

\begin{abstract}

We combined deep Chandra, ROSAT HRI, and XMM-Newton observations of
M87 to study the impact of AGN outbursts on its gaseous
atmosphere. Many X-ray features appear to be a direct result of
repetitive AGN outbursts. In particular, the X-ray cavities around the
jet and counter jet are likely due to the expansion of radio plasma,
while rings of enhanced emission at 14 and 17 kpc are probably shock
fronts associated with outbursts that began $1-2\times10^7$ years
ago. The effects of these shocks are also seen in brightenings within
the prominent X-ray arms. On larger scales, $\sim$50 kpc from the
nucleus, depressions in the surface brightness may be remnants of
earlier outbursts. As suggested for the Perseus cluster (Fabian et
al.), our analysis
of the energetics of the M87 outbursts argues that shocks may be the
most significant channel for AGN energy input into the cooling flow
atmospheres of galaxies, groups, and clusters. For M87, the mean power
driving the shock outburst, $2.4\times 10^{43}$~\ergssec, is three
times greater than the radiative losses from the entire ``cooling
flow''. Thus, even in the absence of other energy inputs, outbursts
every $3\times10^7$ years are sufficient to quench the flow.

\end{abstract}


\section{Introduction}

M87 (NGC4486) provides a unique laboratory to study the interaction
between energy generated by a supermassive black hole and the hot
intracluster medium. Its proximity allowed the disk surrounding its
active nucleus to be resolved by HST providing a measure for the mass
of the central black hole of $3.2\times10^9$~\msun~ (Harms et al. 1993,
Ford et al. 1994, Macchetto et al. 1997). M87's proximity also
provides a unique view of its jet, detected in optical, X-ray, and
radio (e.g., Sparks, Biretta, \& Macchetto 1996, Perlman et al. 2001,
Marshall et al. 2002, Harris et al. 2003).

M87 is the central elliptical in the rich Virgo cluster and is
surrounded by an extensive gaseous atmosphere with a mean temperature
of $\sim2-2.5$ keV (Mathews 1978, Bahcall \& Sarazin 1977, Fabricant
\& Gorenstein 1983, B\"ohringer et al. 2001, Matsushita et
al. 2002). X-ray structure in the gaseous halo of M87 
was first reported by Feigelson et al. (1987) using
Einstein Observatory observations.  Using ROSAT and XMM-Newton
observations, B\"ohringer et al. (1995), Churazov et al. (2001) and
Belsole et al. (2001) discussed the relationship between the observed
X-ray structure and the radio emission.

Like many other optically luminous galaxies at the centers of
clusters, M87 has been considered to be a classic example of a
``cooling flow'' system, where the gas cooling time is relatively
short compared to the age of the system (e.g., Stewart et al. 1984,
Nulsen \& B\"ohringer 1995, B\"ohringer et al. 2001).  However,
observations with XMM-Newton have shown that cooling flows, like that
around M87, deposit cooled gas at much lower rates than expected in
the standard cooling flow model (Fabian 1994; Peterson et al. 2003 and
references therein).  This requires considerable energy input to
compensate for radiative losses.  M87, with its proximity, its active
nucleus, jet, and extensive system of radio lobes, provides an ideal
system for studying the energy input from the AGN to the hot, cooling
gas.

Using radio studies, Owen, Eilek \& Kassim (2000; see also Binney
1999) pioneered the view that the mechanical power produced by the
supermassive black hole at the center of M87 was more than sufficient
to compensate for the energy radiated in X-rays.  Tabor \& Binney
(1993) and Binney \& Tabor (1995) developed models without mass
deposition and included energy injection from the central AGN.  Heinz,
Reynolds \& Begelman (1998; see also Reynolds, Heinz \& Begelman 2001)
modelled shock heating of the IGM by an expanding radio source.
Churazov et al. (2001; see also Kaiser \& Binney 2003, Bruggen 2003,
De Young 2003, Kaiser 2003) argued that the morphology of the X-ray
and radio observations could be explained by radio emitting plasma
bubbles buoyantly rising through the hot X-ray emitting gas. These
buoyant bubbles could uplift the coolest gas and provide energy input
as bubble enthalpy is converted to kinetic energy, then thermalized
into the gas in the bubble wake.

The results outlined above relied primarily on pre-Chandra
observations of M87 that lacked sufficient angular resolution to allow
detailed comparison to the radio structures shown in the Owen et
al. (2000) study. Young, Wilson \& Mundell (2002) used a 37~ksec
Chandra observation to confirm previous structures, as well as to
describe several new features including two nearly spherical ``edges''
at $\sim45''$ and $\sim3'$ that they attribute to activity in the
nucleus associated with jet production. The Chandra images also show
cavities and filaments in the eastern and southwestern X-ray
arms. Young et al. argued that the arms were overpressurized and
multi-temperature. Molendi (2002) used XMM-Newton observations to show
that the X-ray arms required two-temperature models with gas
temperatures in the range $kT\sim0.8-1$ keV and $kT\sim1.6-2.5$ keV. 

In this paper, we combine deep Chandra, ROSAT HRI, and XMM-Newton
observations to study the impact of AGN outbursts on the gaseous
atmosphere around M87 and the interaction of the radio emitting plasma
with the hot gas.  As others have noted, the Chandra, XMM-Newton, and
ROSAT HRI observations show rich structure on many angular
scales. These include knots in the jet, surrounding cavities, a
coincident X-ray and radio bubble just budding from the southeast of
the radio core region, a weak shock, filaments and cavities in the
east and southwest arms, and an arc-like region of enhanced emission
coincident with the outer edge of the southern radio halo. In this
paper, we describe the X-ray observations and discuss the origin of these
features in M87.

\section{Observational Details}

M87 has been well studied at all wavelengths as well as by each new X-ray
mission.  Here, we describe our analysis for Chandra, XMM-Newton, and the
ROSAT HRI. X-ray observations were obtained from the Chandra,
XMM-Newton, or HEASARC archives. While
Chandra provides unprecedented angular resolution, XMM-Newton and
the ROSAT HRI yield large FOV images to study the larger scale
structures around M87.  We adopt a distance for M87 of
16~Mpc (Tonry et al. 2001) which yields a scale of 4.65 kpc per arc
minute.

For all analyses of radial distributions, projections, and spectra,
point sources were detected and then excluded (231 for
Chandra) as was the bright M87 nucleus and jet.

\subsection{Chandra}
\label{sec-chandra_analysis}

M87 was observed with Chandra on 29-30 July 2000 (OBSID 352) and 6-8
July 2002 (OBSID 2707) for 37~ksec and 105~ksec respectively with
ACIS-S at the focus. Details for the July 2000 and July 2002 
observations are given
by Young et al. (2002) and Jordan et al. (2003) respectively. 
The July 2002 observation was used by 
Jordan et al. (2003) to analyze the point source population of M87. 
We applied standard screening to the
event list, omitting ASCA grades 1, 5, and 7, known hot pixels, bad
columns, and chip node boundaries. 

Because the Virgo cluster is both bright and extended, we used the ACIS
S1 chip to monitor the instrument background rate in the energy band
2.5-6.0 keV (see Markevitch 2001). We found significant background
flaring in both observations and the corresponding time intervals were
removed.  The remaining exposure times for OBSID's 2707 and 352 were
87.9 ksec and 30.0 ksec for a total observation time of 117.9 ksec.
For all imaging analyses, we generated images, exposure maps, and
backgrounds separately and then combined the results in sky
coordinates. We normalized the exposure maps between frontside and
backside illuminated CCD's, assuming the emission was characterized by
thermal emission from hot gas with $kT=2$ keV. For spectral analyses,
we extracted spectra separately from each observation and fit the spectra
jointly in XSPEC.  Response matrices and effective areas were averaged
by weighting by the observed X-ray emission.

\subsection{XMM-Newton}

XMM-Newton observed M87 for 57.4~ksec on 19 June 2000. Details for
this observation are given by Bohringer et al. (2001) and Belsole et
al. (2001).  We report on results obtained with the MOS
instrument. Calibrated event lists were generated using SAS v5.3.  The
MOS background was calculated using the blank field data accumulated
over a large number of observations (Lumb et al. 2002) For 
analysis, we used MOS data with patterns in the range 0-12 and the
recommended value of the flag (XMM\_EA). For generating the
temperature map, we use one of the MOS response matrices provided by
the XMM SOC and assume that the same response (corrected for energy
dependent vignetting) is applicable for all regions.  The gas
temperature map was calculated as described by Churazov et al. (1996;
see also Churazov et al. 2003 for this method applied to XMM-Newton
data for the Perseus cluster). An adaptive smoothing also is applied
to the map so that each value of the temperature is calculated using
regions containing $\sim3600$ counts. Comparison of the overall
structure with the results of direct spectral fitting of individual
regions shows good agreement.

\subsection{ROSAT HRI}

M87 was observed seven times with the ROSAT HRI between 1992 and 1997 (ROSAT
sequence numbers rh700214, rh702081, rh704000, rh701712, rh702774,
rh701713, rh702775) for a total observation length of 171.6 ksec. A
detailed description of the ROSAT HRI observations is given by Harris,
Biretta \& Junor (1997) and Harris et al. (2000). We
generated images from each observation using PHA channels 3-9 to
reduce the instrumental background and improve the signal-to-noise for
the diffuse emission. The images were then summed for further analysis.

\section{X-ray Imaging}

X-ray images of the two merged M87 Chandra observations are shown in
Fig.~\ref{fig:bl1sum} and~\ref{fig:adapt}. Fig.~\ref{fig:bl1sum}a
shows the central region at full resolution (1 pixel $= 0.492''$) in
the energy band 0.5-2.5 keV.  Fig.~\ref{fig:bl1sum}b covers a slightly
smaller region and includes 6~cm radio
contours. Fig.~\ref{fig:bl1sum}c labels the features discussed in the
text below.  Fig.~\ref{fig:adapt} is an adaptively smoothed image.  On
the smallest scales, Fig.~\ref{fig:bl1sum} and Fig.~\ref{fig:adapt}
show:

\begin{itemize}
\item The well-known X-ray jet extending $\sim20''$ from the nucleus with
the brightest emission coming from the nucleus and knot A, and showing
emission from at least seven knots (Marshall et al. 2002, Harris et al. 2003).
\item X-ray cavities surrounding the jet (including one surrounding the
location of the unseen counterjet) with an overall extent described
as an ellipse with semi-major and semi-minor axes respectively of
$40''$ and $15''$ oriented along the jet direction (approximately west
northwest) and coincident with the bright radio emission seen at 6~cm
(see Fig.~\ref{fig:inner_cocoon}b).  The radio lobe cavities,
particularly the counterjet cavity, are delineated by bright rims of
X-ray emission.
\item An X-ray cavity (with a radius of about $12''$), delineated by a
bright X-ray rim, coincident with radio emitting plasma ``budding''
from the southeast of the 6 cm radio lobes.
\item An X-ray bright core region, surrounding the 6~cm radio lobes,
with a radius of approximately $50''$ with more pronounced emission to
the north (overexposed region in Fig.~\ref{fig:flatbl1_sm2}).  Young
et al. (2002) suggested that an edge at 3.9~kpc radius arises from sound
waves driven by nuclear activity.  Alternatively, this structure could
originate either as a cold front (see the XMM-Newton temperature map
in Fig.~\ref{fig:xmm_tmap}) or as a sheath of cool material
surrounding the cocoon. The southern edge of the
radio cocoon is bounded by a bright X-ray rim (just south of the
outermost radio contour in Fig.~\ref{fig:inner_cocoon}b).
\item At least four cavities (typical scales of $\sim10''$) extending into
the eastern arm with associated filamentary structures that surround
them.
\end{itemize}

To study the emission on larger scales, we have used the Chandra
ACIS-S2 and S3 CCD's as well as the ROSAT HRI and XMM-NEWTON
observations of M87.  The Chandra large-scale image was prepared by 1)
extracting the 0.5-2.5 keV images from each pointing, 2) generating
``exposure'' maps (accounting for vignetting, quantum efficiency, and
bad pixel/columns/gaps with an assumed $kT=2$ keV thermal spectrum to
normalize the front and back side CCD chips), 3) smoothing (Gaussian
of $1''$) each of the images and exposure maps, 4) dividing each image
by its exposure map and 5) summing the two flat-fielded images. The
resulting Chandra image is shown in Fig.~\ref{fig:flatbl1_sm2}.

To enhance the view of the faint asymmetric structures, we processed
the Chandra, XMM-Newton and ROSAT images to remove the large scale
radial surface brightness gradient. Fig.~\ref{fig:divking} and
Fig.~\ref{fig:rosat} show the relative deviations of the surface
brightness from a radially averaged surface brightness model
($\beta$-model for Chandra and ROSAT and an azimuthal average for
XMM-Newton) i.e. $[{\rm Data}-{\rm Model}]/{\rm Model}$.  These images
show evidence on large scales for buoyant bubbles and energetic
outbursts powered by the supermassive black hole in the M87
nucleus. These features, outside the $1'$ core, are labeled in
{Fig.~\ref{fig:divking}b and Fig.~\ref{fig:rosat}.  The features
include:

\begin{itemize}

\item a nearly azimuthally symmetric ring of emission with a leading
edge at a radius of 14~kpc ($3'$) most prominent to the north and
northwest (see Fig.~\ref{fig:flatbl1_sm2} and Fig.~\ref{fig:divking}),
which we interpret as a weak shock (see below)
\item a second partial ring of enhanced emission, just beyond the
14~kpc ring, at a radius of 17~kpc, most prominent at azimuths of
$0-60$\deg~ from west (see Fig.~\ref{fig:divking}) that we also
interpret as a shock
\item the prominent eastern and southwestern arms that brighten
significantly at approximately the radius of the 14~kpc ring 
\item at radii beyond the 14~kpc ring, the division of each arm into
two filaments. For the southwestern arm, the filaments (labeled {\bf S1}
and {\bf S2} in Fig.~\ref{fig:divking}b) turn east while for the
eastern arm, the two filaments (labeled {\bf E1} and {\bf E2} in
Fig.~\ref{fig:divking}b) turn north.
\item a southern arc, at a radius of approximately 37~kpc ($\sim$8$'$) seen in
Chandra, ROSAT HRI, and XMM-Newton images (Fig~\ref{fig:divking} and
Fig~\ref{fig:rosat}a, b) 
\item on the largest spatial scales, the ROSAT HRI and XMM-Newton
images in Fig.~\ref{fig:rosat}a, b show
depressions in surface brightness to the northeast and southwest and
corresponding excesses to the southeast and northwest (see regions
marked ``Excess'' in  Fig.~\ref{fig:rosat}a, b).
\end{itemize}

\section{XMM-Newton Temperature Map}

The XMM-Newton temperature map with its large field of view and high
signal to noise (compared to Chandra) is shown in
Fig.~\ref{fig:xmm_tmap} and is consistent with the discrete
temperature fits performed by Molendi (2002; see also Belsole et al
2001). The continuous nature of our map clearly shows
some of the large scale structures.  The eastern and southwestern arms
are distinctly cooler than the surrounding gas, although we found no
abundance differences between the arms and the ambient M87 atmosphere.
The temperature map shows, at least as clearly as the surface
brightness map, the clockwise rotation to the east of the southwestern
arm.

Perhaps the most striking feature of the temperature map is the
similarity of the eastern and southwestern arms.  In particular, the
temperatures are similar and at the end of the arms, when they
start to deviate from the approximately linear structure, each bends
clockwise.

\section{Analysis and Interpretation}

We discuss the origin of prominent X-ray features seen in the Chandra,
ROSAT HRI, and XMM-NEWTON images in terms of energetic outbursts from
the supermassive black hole at the M87 nucleus. Previous authors have
provided ages for the eastern and southwestern arms and buoyant
bubbles ($10^7$ years; Churazov et al. 2001) and the outer radio lobes
($10^8$ years; Owen et al. 2000). We note that while we describe the
outer radio lobes as buoyant bubbles, they must be continuously
resupplied with energetic particles as Owen et al. (2000) have
emphasized.

\subsection{Gas Density and Temperature Deprojection}

\label{sec-chandra_deprojection}

To obtain ambient gas density and pressure profiles (see
Fig.~\ref{fig:ring_model}), we deprojected the X-ray surface
brightness and gas temperatures in a sector north of the nucleus at
azimuths (from north) between $-30^\circ$ and $45^\circ$, a region
chosen for its relative absence of asymmetric structure.  The
deprojection is standard (Nulsen et al.~2002), except for the handling
of weights for the outermost region.  Because the outer edge of the
region is well within the cluster, a model was needed to allow for
cluster emission beyond the deprojected volume.  This was done by
assuming that the surrounding gas is isothermal, with a power-law
density profile ($n_{\rm e} \propto r^{-1.09}$, that accurately
matches the profile from $1'-5'$, see Fig.~\ref{fig:ring_model}c).
The outermost ring was assumed to represent emission from the
surrounding gas and its weight was determined accordingly in the
spectral deprojection.

\subsection{X-ray Core and Inner Radio Lobes}

\label{sec-core_lobes}

The inner radio lobes, the cocoon region, originated in an episode of
recent activity. We can use the X-ray observations to estimate the
energetics associated with their formation.  The counterjet cavity is
well-described as an ellipse with semimajor-axes of $15''$ and $19''$
on the plane of the sky (see Fig.~\ref{fig:bl1sum}c).  Using the
innermost gas temperature of 1.65 keV from our deprojection
(Fig.~\ref{fig:ring_model}), and assuming the transverse expansion of
the cavity is subsonic, we estimate that its age is more than
$1.7\times10^6$ years.  The one-sidedness and super-luminal motion of
the jet suggest that its path is close to our line-of-sight.  If the
axis of the cavity makes an angle of 20\deg~ to the line-of-sight
(Biretta, Sparks \& Macchetto 1999), then the head of the cavity is
advancing at $<5 \times10^3$~\kms.  If the cavity is $10^7$ years old,
the average speed of the head of the cavity is mildly supersonic.

If the cavity is prolate and lies in the plane of the sky, its volume
is $\simeq 2.4\times10^{65} \rm\ cm^3$.  Using the pressure for the
innermost ring of our deprojection (centered at a radius of $22''$),
gives $pV = 1.3\times10^{56}$~erg.  If the cavity is relativistic (and
the presence of the synchrotron emitting plasma strongly argues in
favor of this), then its enthalpy is $4 pV$, otherwise it is $2.5 p
V$.  Correcting for the 20\deg~ angle between the line-of-sight and
the axis of the cavity increases the volume of the cavity by a factor
$\sim2.9$ and shifts the center of the cavity to larger radius by the
same factor.  At this position, the gas pressure is 2.6 times smaller,
so that our estimate of the enthalpy is only slightly altered, based
on the assumption that the pressure inside the cavity is uniform (due
to the high sound speed of the relativistic gas). Note that the fact
that the external pressure changes over the extent of the cavity
implies that the configuration must be dynamic -- either parts of the
lobe far from the nucleus (where the external pressure is lower) are
overpressured and thus expanding into the gaseous atmosphere, or the
inner parts, close to the nucleus, are underpressured, and thus
collapsing, or both. The maximum overpressure of a factor of 2.6 would
imply that the head of the cavity is advancing at a Mach number of
1.5.  Since the cavity itself is likely oriented close to the line of
sight, it is quite possible that the assumption of sphericity breaks
down in the inner bins and that the derived pressure is therefore an
overestimate. In either case, a detailed hydrodynamical model is
needed to make a more accurate estimate of the enthalpy, but it is
unlikely to increase over our estimate by more than 50\%, and
certainly not by more than a factor of 2.6.  Doubling the calculated
enthalpy to allow for the jet cavity to the west, the total enthalpy
of these cavities is $\simeq 1.1\times 10^{57}$~erg.  Additional radio
emitting plasma and X-ray cavities surround the jet
cavities. Therefore, given the bubble formation time calculated below,
this material is probably produced in the current outburst as
well. Including this material would at most double the total enthalpy
of the radio plasma from the outburst.

\subsection{Azimuthal Rings}

\subsubsection{The 14~kpc ($3'$) and 17~kpc ($3.75'$) Rings}

The 14~kpc ring is the clearest example of a shock-driven feature in
M87.  Although most prominent to the north and west, it is seen over
nearly 360\deg~ in azimuth centered on the M87 nucleus
(Fig.~\ref{fig:divking}).  As described in
section~\ref{sec-chandra_deprojection}, we measured the surface brightness
and temperature profiles in a sector north of the nucleus.   The
deprojected  gas density profile is shown in
Fig.~\ref{fig:ring_model}, along with the deprojected gas temperature.
If we identify the inner radio lobes with the piston that drives this shock,
since they are small compared to the volume encompassed by the
expanding front, the event that drove the shock may be treated as
instantaneous.  If the shock is expanding into a spherically symmetric
medium, it will become increasingly spherical, regardless of the shape
of the initial outburst.  Thus we model the outburst required to
generate the 14~kpc ring by assuming that energy is injected in a single
event into the atmosphere by the central AGN.  Our estimate of the
outburst energy is
robust to varying details of the actual outburst.
Before the passage of the shock, the gaseous atmosphere
is hydrostatic and isothermal, with a power law density profile
$n_{\rm e} \propto r^{-1.09}$, chosen to match the surface brightness
profile immediately outside the shock.  We derived the parameters of
the outburst by matching calculated surface brightness profiles to
that observed and verifying that the temperature data were consistent
with those calculated. The model that best matches the data is
characterized by an energy deposit of
$8\times10^{57}$~ergs about $10^7$~yrs ago and is shown in
Fig.~\ref{fig:ring_model} (right panel) compared to the observed
surface brightness distribution. The shock is mildly supersonic, with
$M=v/c_{\rm s} = 1.2$ ($v=950$~\kms). A weak temperature enhancement
at the position of the ring (Fig.~\ref{fig:ring_model}) is consistent
with the calculated model (as is the slight temperature reduction
inward of the shock).  Although the temperature evidence is not
strong, the remarkably spherical appearance of this feature is
compelling evidence that it is a weak shock driven by the AGN.  This
description for the formation of the 14~kpc ring is similar to that of
Ruszkowski et al. (2003) in which impulsive energy injection generates
weak shocks as the injected energy forms a cocoon in the galaxy
atmosphere in the Virgo core.  We note that the time estimate,
$1.1\times10^7$ years, for the onset of the activity that gave rise to
the 14~kpc ring is robust and the simple model gives this age to an
accuracy of about 10\%.

A second partial ring is seen to the west at a radius of 17~kpc
($3.75'$) extending over $\sim60$\deg~ in azimuth (see
Fig.~\ref{fig:divking}c).  To form this surface brightness
enhancement, a disturbance traveling at the sound speed would have
originated approximately $4\times10^6$ years before the event that
created the 14~kpc ring.  The amplitude of the 17~kpc ring is
comparable to that of the 14~kpc ring and therefore would require a
similar amount of injected energy.  The timescale between the two
events lies well within the range, $10^5$--$10^8$~yrs, expected for
repetitive AGN outbursts.

In our shock model for the 14~kpc and 17~kpc rings, one or both may be
associated with the inflation of the inner cocoon.  This is consistent
with radio source models in which the radio outburst commences with
rapid expansion, driving a shock.  As the expansion slows, the shock
separates from the driver, weakens and becomes more spherical (the
``sonic boom phase'' in Reynolds, Heinz, \& Begelman 2001).  The
current energy input by the jet is estimated to be $10^{44}$~\ergssec~
by Bicknell \& Begelman (1996). Owen et al. (2000) used the radio
observations to derive a minimum value of the instantaneous energy
input by the jet of a few $\times10^{44}$~\ergssec. The energy input is
more than sufficient to power the lobes and generate a roughly
spherical pulse.

The spherical shock model provides an estimate of the total energy in
the outburst. This energy is significantly greater (a factor of 3 or
more) than the enthalpy of the cavities also created by the AGN
outburst (see section~5.2). This suggests that shocks
may be the most significant channel of AGN energy input 
into ``cooling flow'' atmospheres in early type galaxies,
groups, and galaxy clusters. Fabian et al. (2003) find similar shocks
and ``edges'' 
around NGC1275 in the Perseus cluster. They
also argue that the energy deposited as these features dissipate
can compensate for the energy radiated by the cooling gas in the 
inner 50~kpc core. For M87, the mean power of the shock
outburst averaged over the past $10^7$ years is $2.4\times10^{43}{\rm\
erg\ s^{-1}}$. Within 70~kpc, the radiative loss from the ``cooling
flow'' is $\sim10^{43}$\ergssec~ (corresponding to 10~\msun~yr$^{-1}$;
Stewart et al. 1984).  Thus, one such outburst every $3\times10^7$
years is sufficient to quench the cooling flow in the absence of any
other heat source.  In addition to the energy input from the outburst
we have modelled, energy is also being supplied by the buoyantly
rising features seen in the radio maps (Churazov et al. 2002).

\subsubsection{Southern 37~kpc ($8'$) Arc}

South of the nucleus at a radius of 37~kpc ($8'$), a surface
brightness enhancement appears as an arc or partial ring in
observations by Chandra, ROSAT HRI, and XMM-Newton (see
Fig.~\ref{fig:divking} and Fig.~\ref{fig:rosat}a, b). This partial
ring extends over an azimuth of at least 45\deg. Statistically, its
significance is shown in the radial profile in
Fig.~\ref{fig:south_profile} made from the Chandra image. As
Fig.~\ref{fig:rosat} shows, the southern 37~kpc arc lies just outside
the large scale radio lobes characterized by Owen et al. (2000) as the
oldest ($10^8$ yrs) structures in M87.  In the model by Churazov et
al. (2001), these radio lobes were originally buoyant bubbles that
have risen in the gaseous atmosphere surrounding M87 and are now thin
disks (``pancakes'') seen in projection. Each $10^8$ yr old bubble is
partially surrounded by gas that has been displaced by the bubble's
rise.  This gas could be material piled up on the edge of the bubble
during its expansion or alternatively, gas that was uplifted by the
bubble and is draining off the bubble along its sides.  The XMM-Newton
spectra show that the gas associated with this feature has a similar
temperature ($kT\approx2.5$ keV) but high abundance ($1.2\pm0.3$ of
solar for an APEC model) compared to other gas at the same radius
(0.4-0.5 of solar).

Rims of cool gas are a common feature of cavities created by radio
lobes (e.g.\ Finoguenov \& Jones 2001, McNamara et al.\ 2000, Fabian
et al.\ 2002, Blanton et al.\ 2001).  In M87 we see rims most clearly
in the bright edge of the counterjet cavity, the southeastern ``bud'',
and the southern edge of the outer radio lobes. The presence of bright
rims over such a wide range of scales illustrates the ability of the
radio plasma to exclude hot gas and attests to their surprising
stability (Nulsen et al. 2002).

At the western end of the 37~kpc arc, a brightness enhancement is
seen in both the XMM-Newton and ROSAT HRI images
(Fig.~\ref{fig:rosat}a, b). This feature is extended from northwest to
southeast ($30''$, 2.3~kpc, in length) and is likely associated with a
Virgo cluster member, the E2 galaxy NGC4478 (RA=12:30:17.4,
DEC=+12:19:43) that is coincident with the X-ray feature.  The X-ray
luminosity of the galaxy is $7\times10^{38}$ \ergssec~ and its optical
absolute magnitude is $M_B=-18.8$ (for a distance of 16~Mpc). The
X-ray and optical luminosities are consistent with the $L_x-M_B$
correlation for the emission from hot gas in early type galaxies
(e.g., Forman, Jones \& Tucker 1985).  David, Forman \& Jones (1991)
showed that early type galaxies with $M_B=-19$ would be transitioning
from atmospheres with partial to total subsonic winds, assuming a
supernova rate of 0.15 SN Ia per $10^{10} L_\odot$ per century (van
den Bergh, Maclure \& Evans 1987). If the supernova rate in NGC4478 is
slightly less than this assumed value or if some fraction of the
supernova energy is not transferred to the hot atmosphere, then
NGC4478 could maintain a barely stable atmosphere. Such an atmosphere
could be ram pressure stripped in the dense Virgo core, producing the
extended X-ray emission seen in the ROSAT HRI and XMM-Newton
images. Thus, most likely, we are observing the gaseous halo of an
elliptical galaxy being stripped by the atmosphere around M87 and the
feature is probably not associated directly with the southern arc.

\subsubsection{Large Scale Surface Brightness Asymmetries}

On the largest scales, the XMM-Newton and ROSAT images
(Fig.~\ref{fig:rosat}a, b) show surface brightness enhancements and
depressions outside the outer radio lobes and beyond the 37~kpc arc. 
In particular, in 30\deg~ wide sectors at a radius of $50\pm5$~kpc, the
surface brightness in the ROSAT HRI has a maximum 18\% above the
mean in the northwest and a minimum 10\% below the mean in the southwest. 
At somewhat smaller radii, Fig.~\ref{fig:rosat} shows higher surface
brightness to the southeast, just beyond the outer radio lobe. 

Some of these asymmetries could arise from an elliptical gravitational
potential or they could result from subtracting an azimuthally
symmetric model for the surface brightness from an elliptical
distribution. Alternatively, such an asymmetrical distribution could arise from
``ghost cavities'' (e.g., Ensslin 1999) of relativistic plasma,
produced by earlier epochs of AGN activity, but no longer emitting at
observable radio frequencies.  The outer radio lobes are approximately
$10^8$ years old and, as Owen et al. (2000) argued, their detection
requires ongoing energy injection. Any older cavities, at larger
radii, may no longer have a connection to the nucleus and, hence,
would no longer be detectable at radio wavelengths.

\subsection{A Budding Bubble}

In most systems, X-ray cavities or radio plasma bubbles generally are
consistent with having been directly filled by an active jet and are
aligned with the jet axis or lie on opposite sides of the active
nucleus.  Clear examples of such alignments include a wide variety of
systems ranging from galaxies with very modest gaseous atmospheres
like Cen A (Kraft et al. 2003) and M84 (Finoguenov \& Jones (2001) to
rich luminous clusters including Hydra~A (McNamara et al. 2000, David
et al. 2001, Nulsen et al. 2002), and Perseus (B\"ohringer et
al. 1996, Fabian et al. 2002).

While many of the inner X-ray structures in M87 are clearly aligned
along the direction of the M87 jet axis, the southeastern bubble
(extending from $30''$ to $45''$ from the M87 nucleus) is an exception
(see Fig.~\ref{fig:bl1sum}).  This bubble corresponds precisely to a
radio feature and appears as a ``bud'' emanating from the southeast of
the bright radio core.  Fig.~\ref{fig:bud}b 
shows the X-ray image with the 6~cm radio
contours superposed (Hines, Owen \& Eilek 1989).  The surrounding
X-ray emission traces the outline of the outer radio contour and shows
that the X-ray cavity is filled with radio emitting plasma.

The ``bud'' emanates from the inner radio cocoon (just south of the
cavity corresponding to the counterjet), almost perpendicular to the
axis defined by the jet. While its origin is likely associated with an
outburst from the active nucleus and an episode of energy injection
into the inner cocoon, the location of the bubble may mark a
magnetically weak region of the inner cocoon (the counterjet cavity)
with buoyancy forces driving the bubble
perpendicular to the dominant axis of recent activity.

If we assume that the rise of the budding bubble (radius $r =
11''$, 0.85~kpc) is governed by buoyancy and limited by the drag of
M87's hot gaseous atmosphere, we can estimate the formation time,
$\tau_{\rm bubble}$, as the time for the bubble to rise through its
own diameter.  If $R$ is the distance from the cluster center to the
current position of the bubble and $M(R)$ is the total gravitating
mass within $R$, then
\begin{equation}
\tau_{\rm bubble} \simeq 2 R \sqrt{C_{\rm W} r \over GM(R)}
\simeq 4 \times 10^6 {\rm\ y}, \label{eq:risetime}
\end{equation}
where we have used $R = 34''$ ($2.6\,{\rm kpc}$), the projected
distance from the cluster center, and $M(R) = 1.4\times
10^{11}\,M_{\odot}$ (C\^ot\'e et al.\ 2001).  $C_{W} \sim 0.5$ is the
drag coefficient for a roughly spherical bubble.  Since the actual
distance to the cluster center almost certainly exceeds the
projected distance, this gives a lower limit for the rise time.
Furthermore, according to (\ref{eq:risetime}), the speed of the
bubble, $383\rm\ km\ s^{-1}$, exceeds half of the sound speed and so
is overestimated.  Thus its rise time is underestimated, even if the
budding bubble lies in the plane of the sky.

During its rapid initial expansion, the boundary of the bubble will
generally be stable.  As a result, the motion of the bubble boundary
generally needs to be subsonic before a bubble even starts to form.
This adds a further delay to $\tau_{\rm bubble}$ after the outburst,
but before the bubble is formed.  If we do associate the budding bubble
with an energetic nuclear event, then the constraints on its formation
timescale make it quite reasonable to associate it with the current
outburst (associated with the jet) that commenced about $10^7$ years
ago.

\subsection{Buoyant Bubbles and the Structure of the Eastern 
and Southwestern Arms}
		
The most striking X-ray features in M87 are the two arms that extend
east and southwest from the inner lobe region. These also are  seen in
the 90~cm image (see Fig.~\ref{fig:overlay} for a composite
X-ray-radio view of M87). Previous spectroscopic studies of the arms
have utilized the XMM-Newton observations (Belsole et al. 2001,
Molendi 2002). They find that the arms are cool and portions are
poorly fit by single temperature components. Our Chandra results agree
with these previous analyses, as does the XMM-Newton temperature map
(Fig.~\ref{fig:xmm_tmap}). We find that the arms require at least two
components (with variable abundances, VMEKAL or VAPEC) with the low
and high temperature components in the range 1-1.5 keV and 2-2.7 keV
respectively.  Although the two arms are likely related to the same
outburst, we discuss each separately.

\subsubsection{Eastern Arm}

The eastern X-ray and radio arm begins at the eastern edge of the
inner radio cocoon, but its appearance is much more amorphous than
that of the southwestern arm (see Fig.~\ref{fig:bl1sum},
\ref{fig:flatbl1_sm2}, and \ref{fig:divking}). At the base of the
filament (Fig.~\ref{fig:bl1sum} and \ref{fig:adapt}) are at least four
bubbles with sizes comparable to that of the ``bud'' discussed above
and streamers of gas bounding these buoyantly rising bubbles. Typical
bubble sizes are $\sim10''$ (0.8~kpc) in radius and are reminiscent of
the ``effervescent'' heating described by Begelman (2003).
Fig.~\ref{fig:bubble_proj} shows a projection across one of these
``effervescent'' bubbles $1.25'$ (5.8~kpc) east of the M87 nucleus
(labeled ``bubble'' in Fig.~\ref{fig:bl1sum}c).

The temperature structure (Fig.~\ref{fig:xmm_tmap}) of the eastern arm
shows X-ray features that are consistent with cool material uplifted
by a rising torus (Churazov et al. 2001). First, the largest
concentration of the coolest gas lies midway along the eastern arm
($1'-2'$ from M87's nucleus). Second, the cool gas column in the
eastern arm narrows at the edge of the radio torus closest to the M87
nucleus and then broadens within the torus (labeled ``Uplifted Gas''
in Fig.~\ref{fig:divking}b), just as one might expect for gas uplifted
by a buoyant toroidal plasma bubble (see Fig.~\ref{fig:overlay} and
Fig.~3 and~4 in Churazov et al. 2001).

A projection along the arm, Fig.~\ref{fig:eastern_arm_proj}, shows a
25\% brightening at the radial distance of the 14~kpc ring. A similar
brightening occurs at about the same angular distance on the
southwestern arm. While the feature in the southwestern arm is partially
obscured by the change from the ACIS S3 to S2 chip in the Chandra
image, it is clearly seen in both the ROSAT HRI and XMM-Newton images
(Fig.~\ref{fig:rosat}).  If this brightening is associated with the
passage of the same shock that produced the ring, then this arm (and
the southwestern arm as well) must lie close to the plane of the sky.

If this brightening does arise from the passage of the shock, it is
likely that the so called ``radio ear'', the vortex-like structure
that forms the end of the bright eastern radio filament (see
Fig.~\ref{fig:overlay}), falls between the shocks associated with the
14~kpc and 17~kpc rings.  This could alternatively explain the flat,
ring-like appearance of this radio feature, since passage of a shock
through a bubble of relativistic plasma embedded in a background of
cold thermal material will induce strong vorticity in the plasma,
turning it into a ring-like structure (Ensslin \& Bruggen
2002). Combined with the effect of vorticity creation in buoyantly
rising bubbles described by Churazov et al. (2001), this could account
for the rather filamentary appearance of this feature.

At the end of the eastern arm ($\sim3'$ east of the M87 nucleus), the
X-ray image (Fig.~\ref{fig:divking}) shows an almost circular
enhancement (radius of $1'$ centered at RA=12:31:05.397
DEC=+12:25:10.01) extending to the north (beyond the northern ``ear''
of the radio emitting torus). This circular feature is bounded on
three sides by X-ray enhancements (see Fig.~\ref{fig:divking}) which
originate at the eastern arm and it is bounded to the northwest by a
pair of radio arcs (best seen in the 90~cm image; see
Fig.~\ref{fig:overlay}).  The X-ray temperature of this circular
region is intermediate in temperature (1.8-1.9 keV) as seen in
Fig.~\ref{fig:xmm_tmap} and is
comparable to that of the end of southeastern arm (as it swings to the
east). The two enhancements, labeled {\bf E1} and {\bf E2} in
Fig.~\ref{fig:divking}c), which bound the circular region, appear
similar to the two filaments into which the southwestern arm divides
(see below). We suggest that the outer portions of the eastern arm are
similar to the southwestern arm, but seen from a different
orientation.

\subsubsection{Southwestern Arm}

The southwestern X-ray arm originates (see Fig.~\ref{fig:inner_cocoon}
and Fig.~\ref{fig:adapt}) as a narrow filament of width approximately
$10''$ (0.8~kpc) at its narrowest when it exits from the bright inner
core (at a distance of $50''$, 3.9~kpc from the nucleus). The filament
extends in an almost straight line to the southwest for $\sim 2'$
(9.3~kpc). As seen in Fig.~\ref{fig:overlay}, over this distance it
appears uncorrelated with the radio filament that extends in
approximately the same direction. At a distance of about $3.4'$
(15.8~kpc), the X-ray filament bifurcates (the two sections are
labeled {\bf S1} and {\bf S2} in Fig.~\ref{fig:divking}c).  and the
correspondence between the radio plasma and X-ray gas becomes more
direct.  The brightest radio emission lies between the two X-ray arms
as they both rotate clockwise in the plane of the sky and eventually
turn due east.

Young et al. (2002) suggested that the arms are
overpressurized. Assuming the southwestern arm is a cylinder lying in
the plane of the sky, we find that the pressure in the arm is roughly
twice that of the hotter ambient gas. We found no elemental abundance
differences that could explain the surface brightness enhancement in
the arms. However, the pressure difference seems unphysical, since the
time for the pressure to come to equilibrium (the sound crossing time
of the arm) is short compared to the sound travel time along the
length of the narrow arm.  Therefore, the axis of the arm would need
to make a sufficiently small angle with our line-of-sight (roughly
15\deg) to make its projected path length about 4 times its width, in
order to account for its high emission measure. This seems unlikely,
given that the brightening along the arm is at a comparable radius to
that of the eastern arm and both are at a radius similar to that of
the 14~kpc ($3'$) ring.  However, in
computing the overpressure of the arm, we did not include any
contribution to the external pressure from non-thermal particles or
from magnetic field which could contribute to the confinement. Magnetic
tension could also serve to confine the gas in the southwestern arm
since the radio emission from the southwestern arm appears to spiral
around the X-ray gas (see Fig.~\ref{fig:overlay}).
Alternatively, the southwestern arm may not
be formed by rising bubbles as seen in the eastern arm, but is instead
a thin sheath of gas exterior to a large plasma bubble related to the
southwestern radio arm.  This is consistent with a roughly 20\%
increase in surface brightness from west to east across the arm. 

\section{Conclusions}

We have presented a discussion of several of the remarkable structures
seen in the Chandra, XMM-Newton, and ROSAT HRI observations of
M87. Many of these, particularly the bubbles emanating from the
central region, the nearly circular rings of enhanced emission at
14~kpc and 17~kpc, and the brightening of the X-ray arms at these
radii, can be attributed to AGN outbursts.  The 14~kpc and 17~kpc
rings,  similar to the ``ripples'' seen in the 
Perseus cluster (Fabian et al. 2003),  can be interpreted as shock 
waves driven by the current outburst that began about $10^7$ years ago.
The outburst also inflated the
inner radio lobes (and cocoon). Outbursts like those that produced
these shocks can quench the M87 cooling flow, if they occur
approximately every $3\times10^7$ years. Since the enthalpy associated
with the inner cavities, produced by the outburst, is only 30\% of the
energy of the outburst, shock heating is probably the dominant heating
mechanism for the gas in the inner regions of ``cooling flow''
systems.

At larger radii, we see highly enriched gas along the outer edge of
the southern radio lobe (the 37~kpc arc). Asymmetric gas
distributions at radii of $\sim$50~kpc may be evidence for older
($>10^8$ years) outbursts as are the outer radio lobes.

The hot X-ray emitting gas contains reflections of previous episodes
of AGN activity in the form of bubbles and their bright rims, shocks,
and buoyantly uplifted gas structures.  With the detailed observations
at X-ray and radio wavelengths of M87, we can probe the interaction
between the central AGN, the relativistic plasma, and the X-ray
gas. We are beginning to understand the cyclic heating of the X-ray
gas and the energy transfer mechanisms between the central
supermassive black hole and the hot gaseous atmosphere 
that surrounds central cluster galaxies.

\acknowledgements

We acknowledge stimulating discussions with D. Harris, L. David,
M. Begelman, and R., Sunyaev.  This work was supported by NASA
contracts NAS8-38248, NAS8-01130, NAS8-03060, the Chandra Science
Center, the Smithsonian Institution, and MPI f\"{u}r Astrophysik.

\clearpage

\begin{figure*}[ht]
  \centerline{\includegraphics[width=0.52\linewidth]{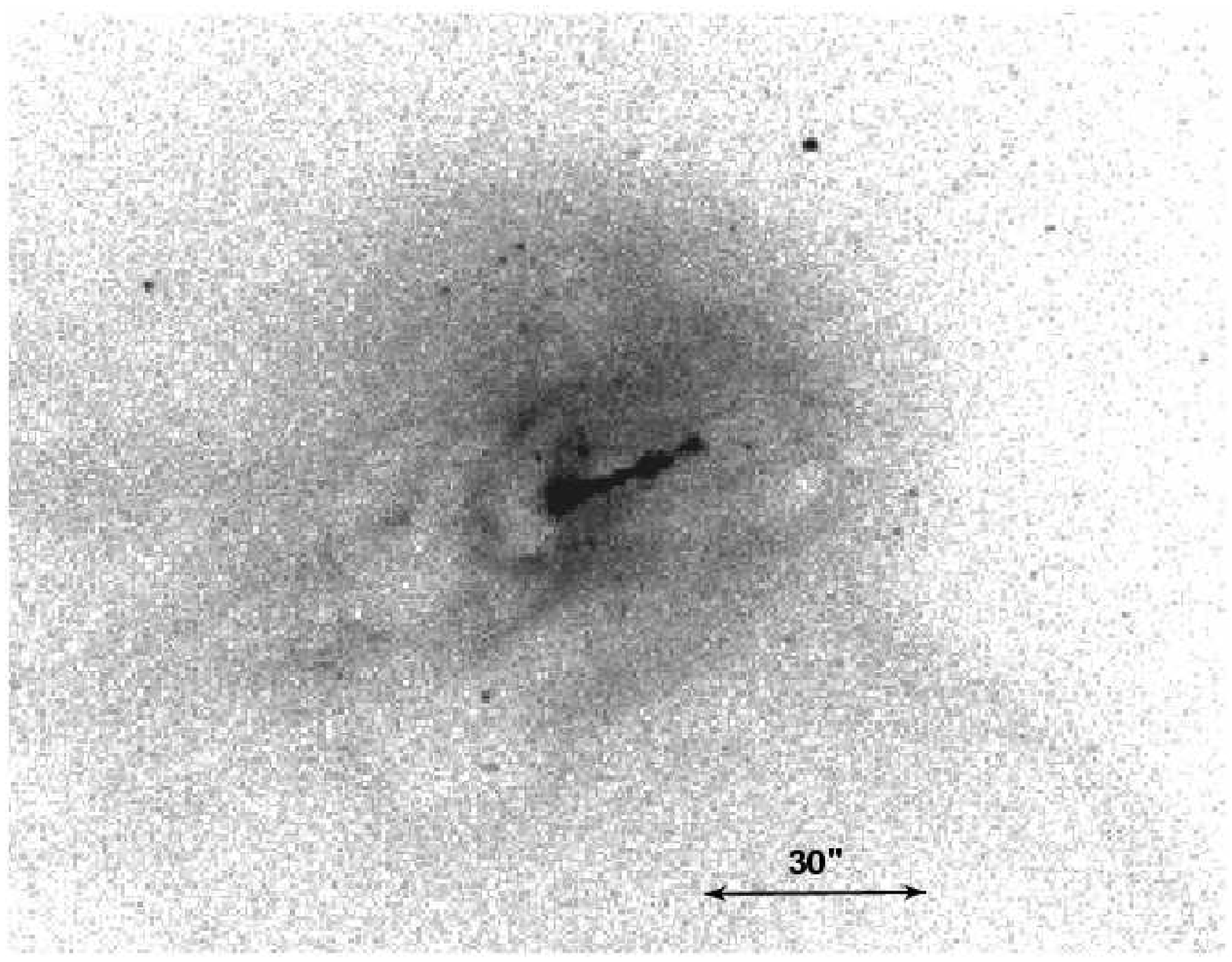}
  \includegraphics[width=0.44\linewidth]{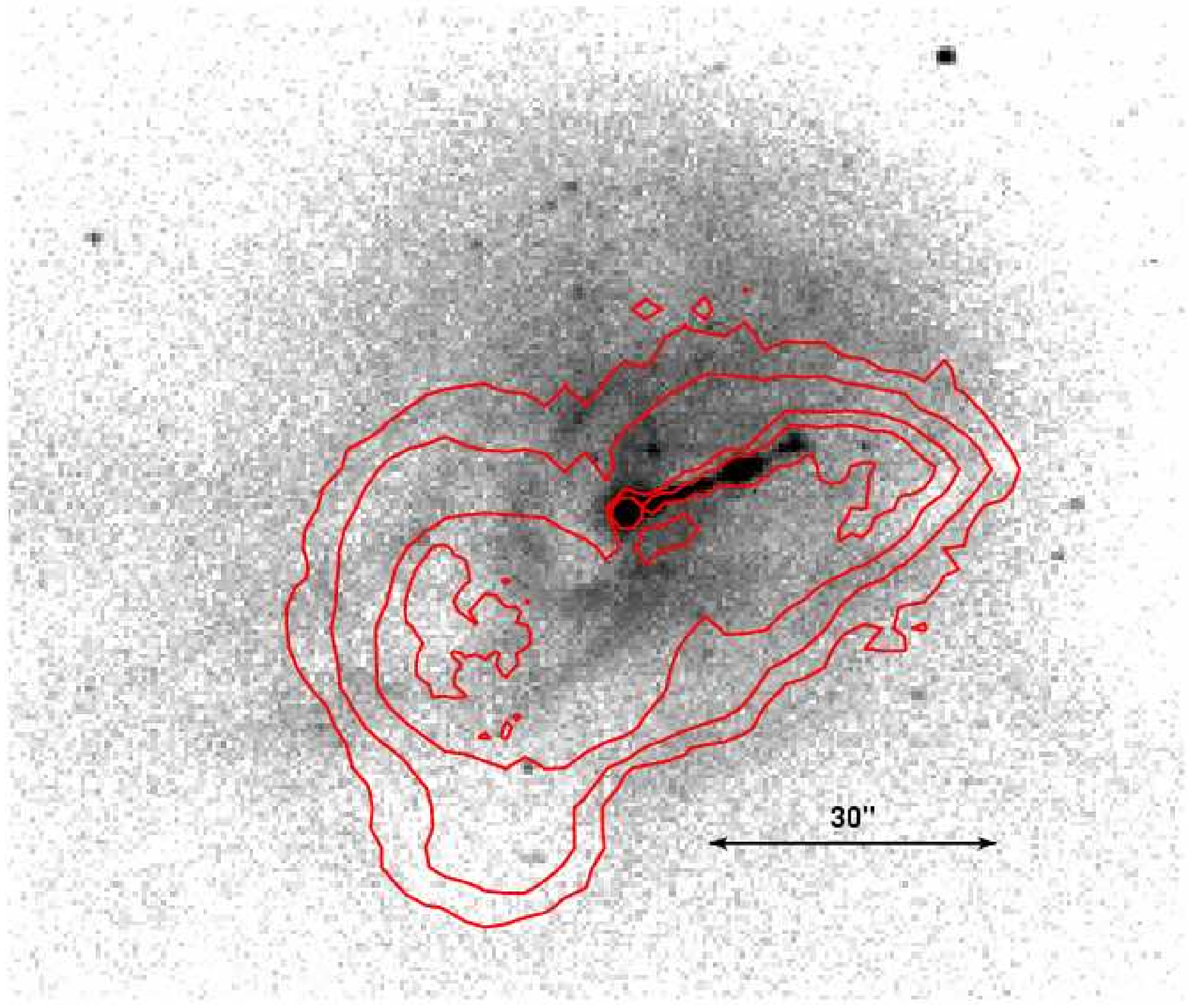}}
  \centerline{\includegraphics[width=0.75\linewidth]{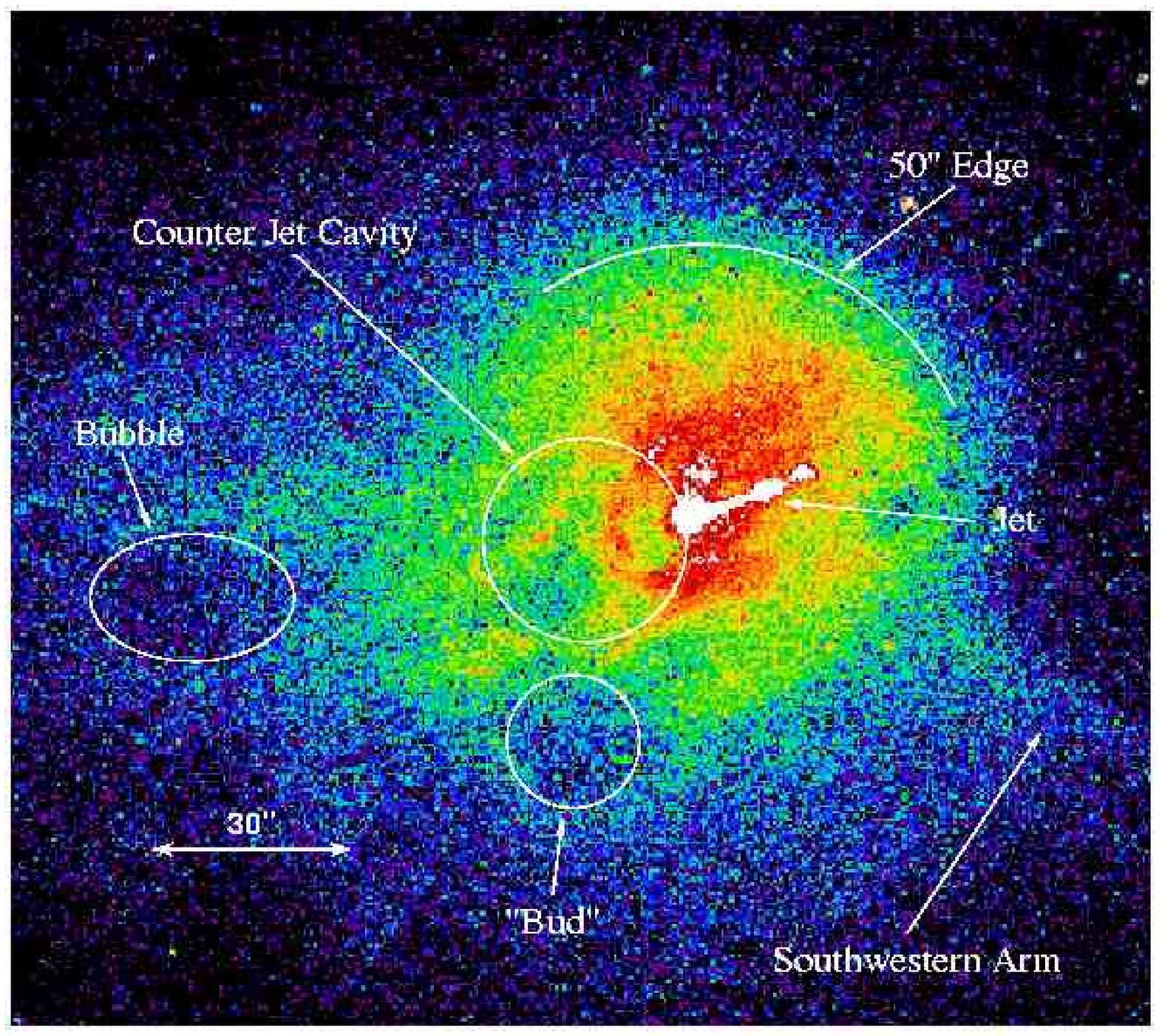}}
  \caption{(a) The top left panel shows the central region of M87 as
  seen by the Chandra ACIS-S detector in the energy band 0.5 to 2.5
  keV with a Gaussian smoothing of 1 pixel = 0.492\arcsec. Several
  cavities are seen in the counter jet direction as is the beginning
  of the large scale eastern arm. (b) The top right panel shows the
  same image at a slightly larger scale with contours (1, 5, 20,
  45$\times10^{-5}$ JY per $0.1\times0.1''$ pixel) from the 6~cm radio
  observations of Hines et al. (1989) superposed on the X-ray image.
  A ``bud'' of radio emitting plasma fills the X-ray cavity to the
  south-east. The region referred to as the cocoon in the text is the
  radio emitting plasma filled region defined by the radio contours
  (excluding the ``bud''). X-ray filaments surround the bubbles in the
  eastern arm.  The bottom panel (c) identifies features
  discussed in the text. 
}
\label{fig:bl1sum}
\label{fig:inner_cocoon}
\label{fig:center}
\label{fig:bud}
\end{figure*}

\begin{figure}
  \centerline{\includegraphics[width=0.95\linewidth,bb=68 30 525 385,
  clip]{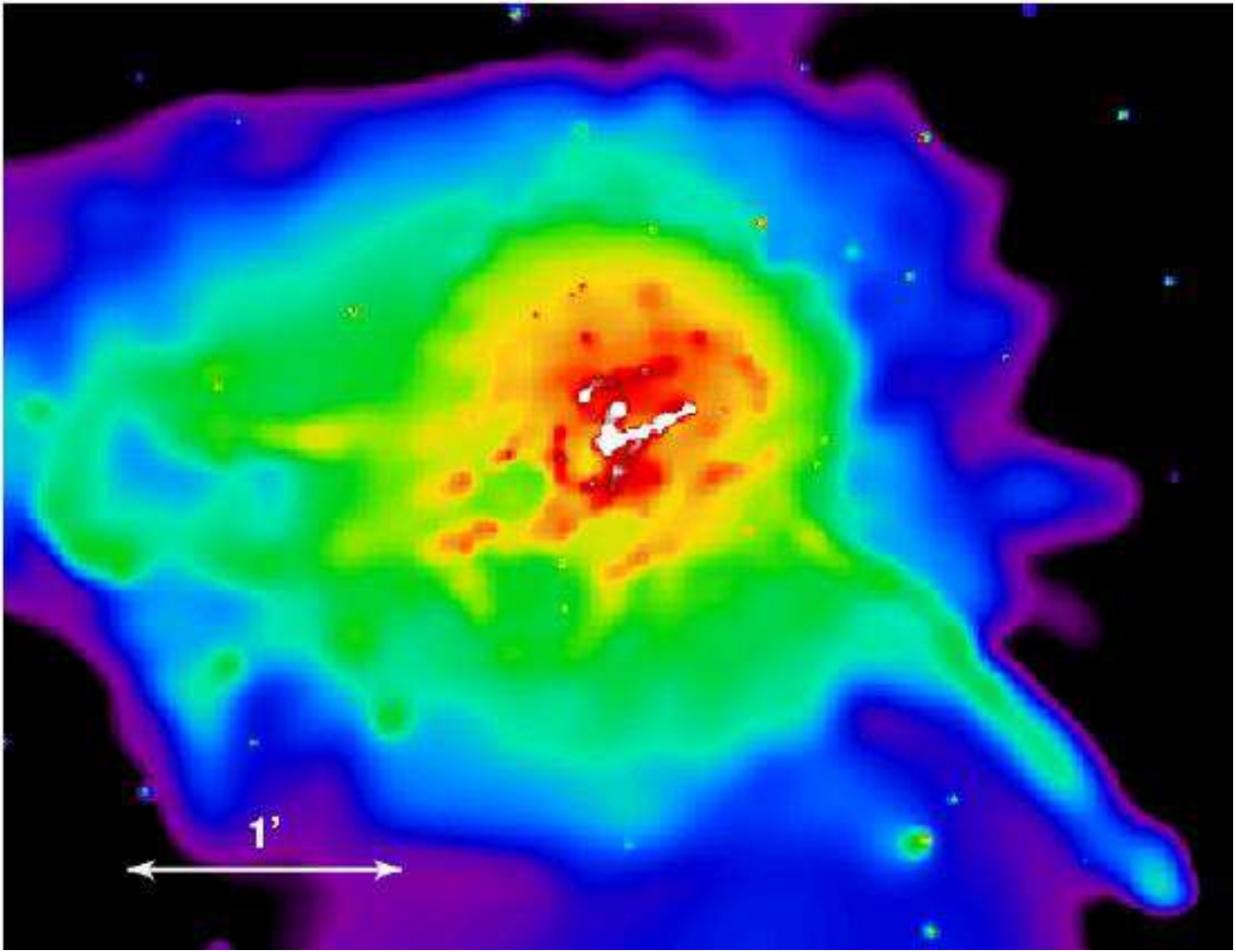}}
  \caption{An adaptively smoothed Chandra image (minimum significance
  $4\sigma$) of the central region of M87. Prominent features include
  the narrow southwestern arm, the bright $50''$ radius inner core
  (yellow) which shows an  especially sharp jump in surface brightness
  along the northern edge, 
  and multiple bubbles and surrounding filaments in the core
  that form the base of the eastern arm.}
\label{fig:adapt}
\end{figure}

\begin{figure}
  \centerline{
  \includegraphics[width=0.95\linewidth]{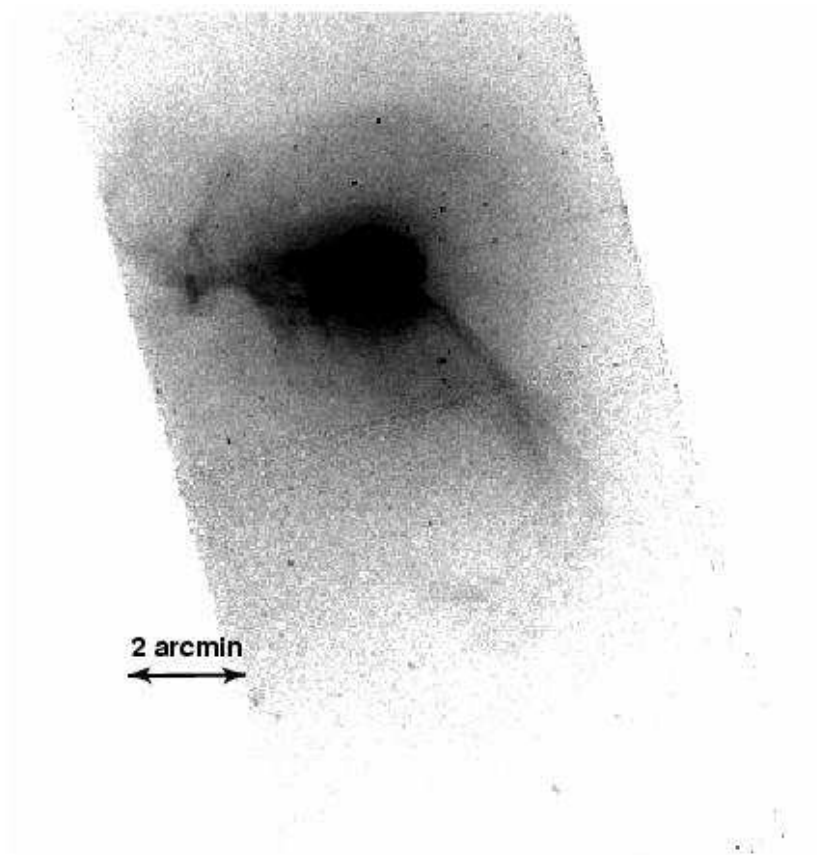}}
  \caption{The merged, flat-fielded Chandra image in the energy band
  0.5-2.5 keV generated by summing the two pointings after
  flat-fielding and smoothing each separately (see text for
  details). The prominent eastern and southwestern arms are apparent
  as is the surface brightness enhancement with an outer edge at $3'$
  radius from the nucleus.}
\label{fig:flatbl1_sm2}
\end{figure}

\begin{figure}
\centerline{\includegraphics[width=0.50\linewidth,bb=120 40 475 530,clip]{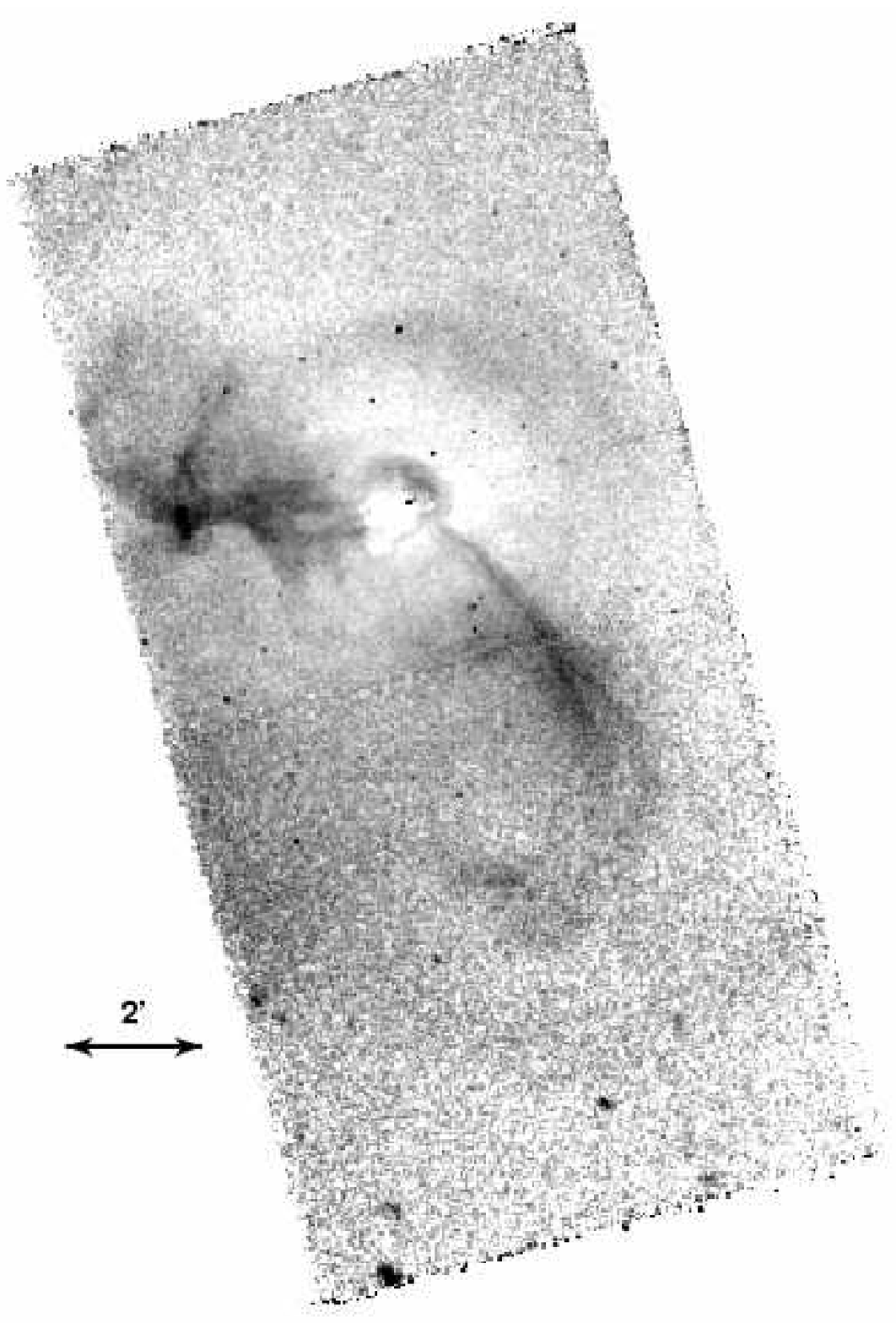}
  \includegraphics[width=0.50\linewidth,bb=155 25 490 480,clip]{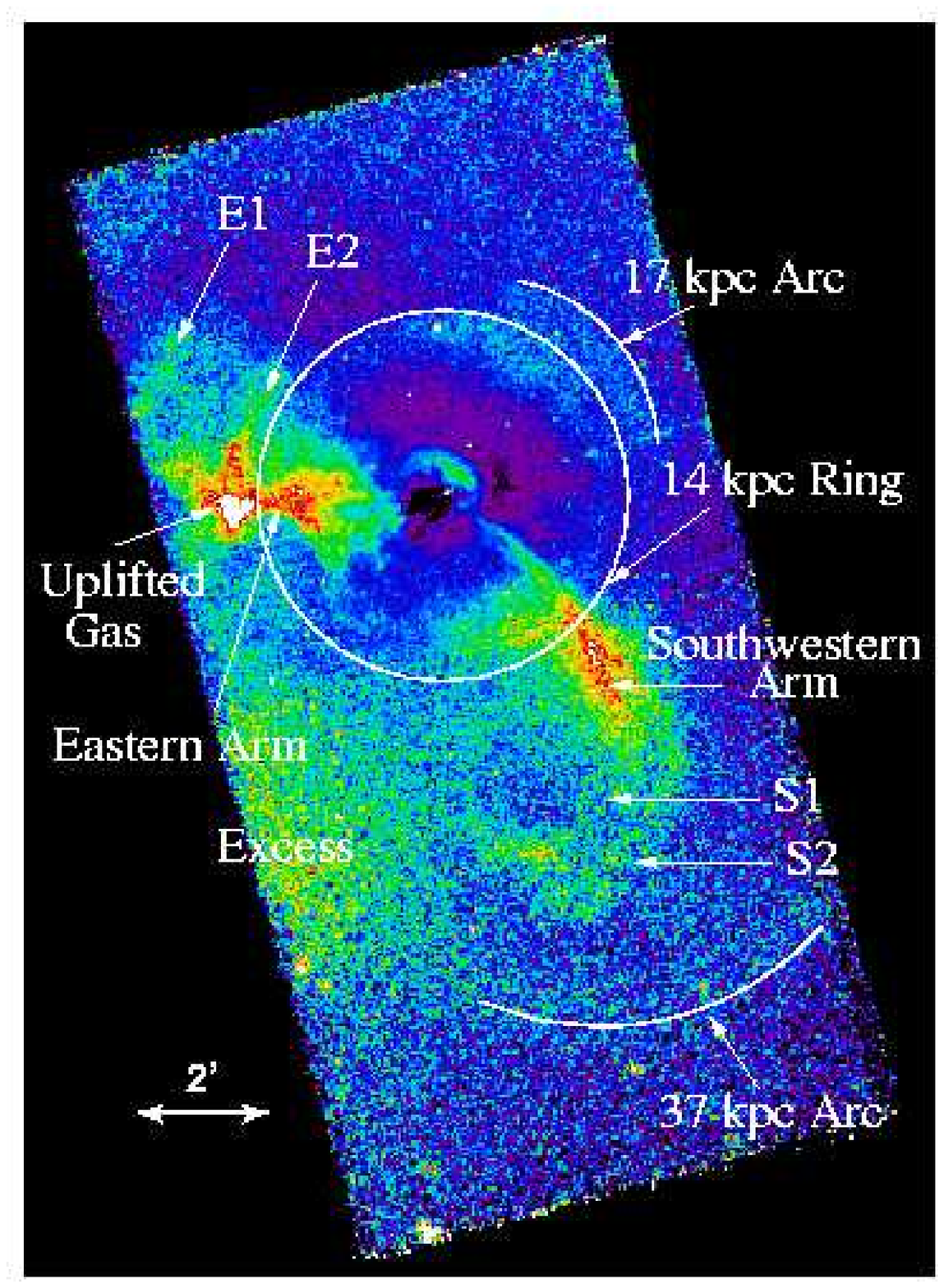} }
  \caption{The Chandra image (0.5-2.5 keV) processed as described in
  the text to remove the large scale radial surface brightness
  gradient. Many faint features are seen including 1) the bifurcation
  of the eastern and southwestern arms, 2) the brightening at the
  eastern and southwestern arms, 3) the 14~kpc ($3'$) ring, 4) the
  17~kpc ($3.75'$) arc, and 5) the faint southern 37~kpc ($8'$)
  arc. The features are labeled in the color image. The circle and arc
  for the 14~kpc and 17~kpc ring and arc are drawn at their outer
  extents. {\bf E1, E2} and {\bf S1, S2} identify the extensions of
  the eastern and southwestern arms after each has divided.  The
  uplifted gas core (white in the color image) that lies at the end of
  the eastern arm at the (projected) center of the radio torus is
  labeled. }
\label{fig:divking}
\end{figure}

\begin{figure*}[t]
\centerline{\includegraphics[width=0.6\linewidth]{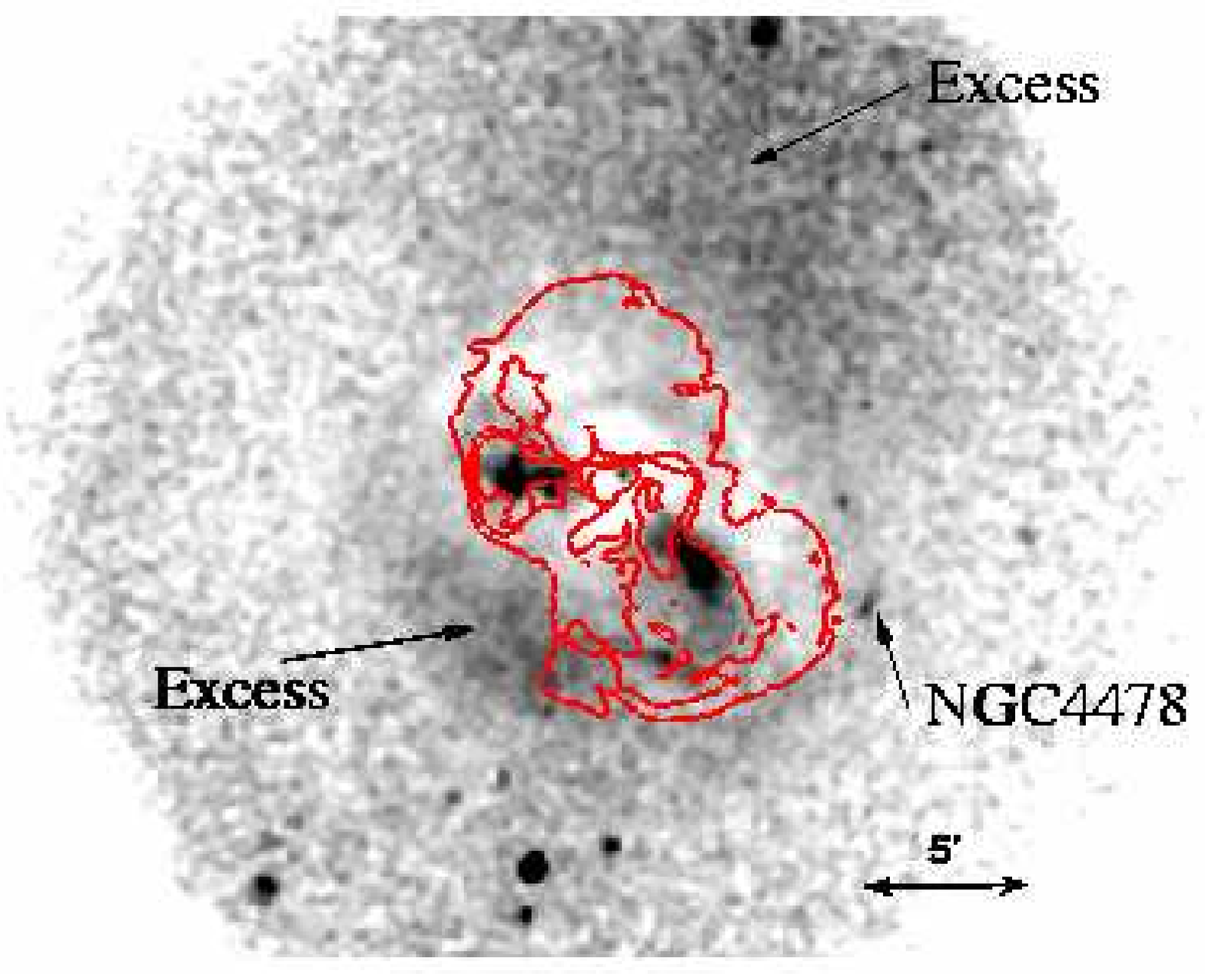}}
\centerline{\includegraphics[width=0.9\linewidth]{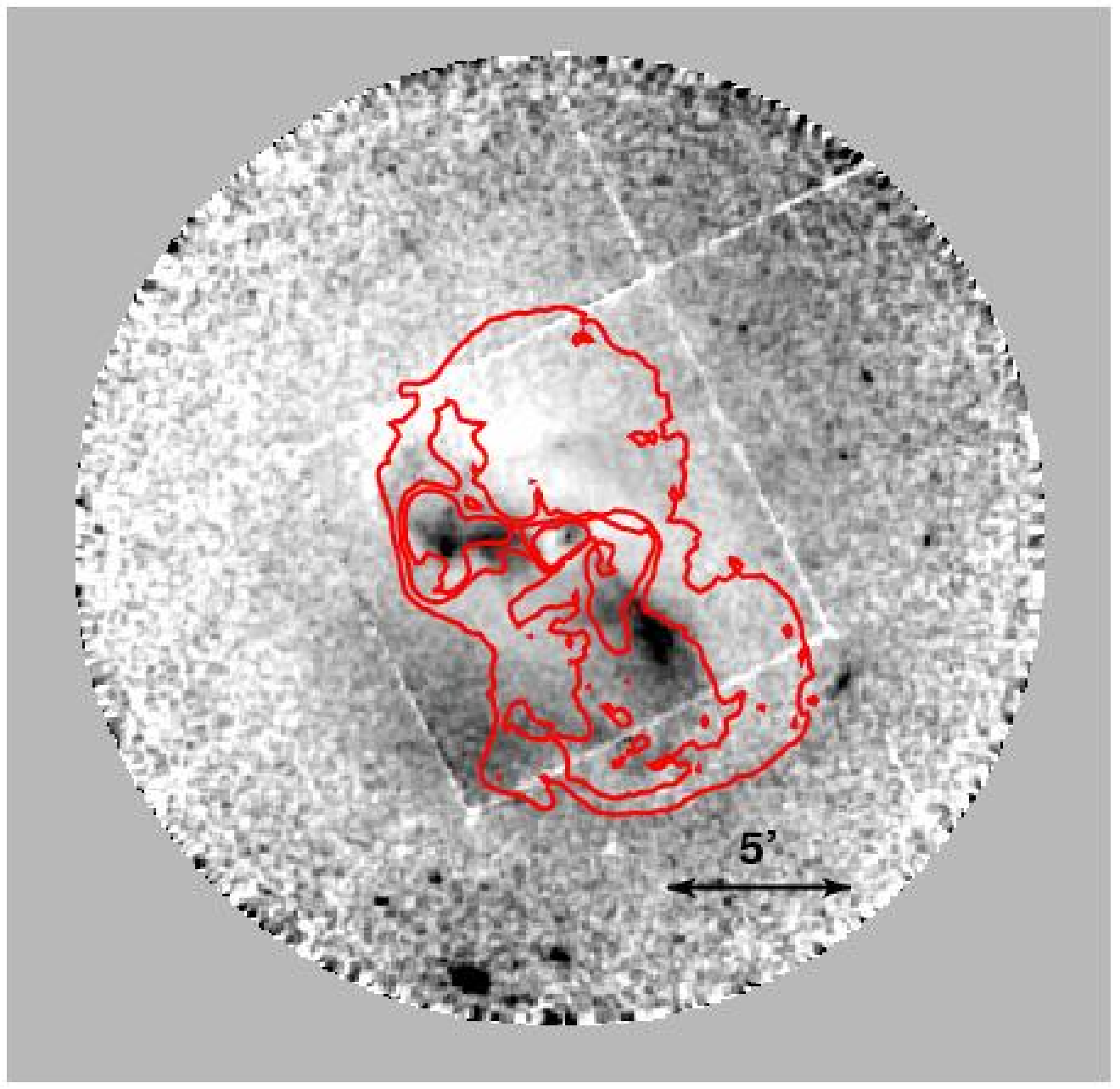}
}
\caption{(a - top) The ROSAT HRI and (b - bottom) the XMM-Newton MOS1
+ MOS2 images processed to remove the steep radial surface brightness
gradient. Both figures show asymmetric emission on large scales
(beyond the outer radio lobes). The diffuse emission (labeled
``Excess'' in the ROSAT HRI image) is bright to the east of the southern
radio lobe, as well as to the north of the northern lobe.  Asymmetric
gas distributions at radii of $\sim$50~kpc may be evidence for older
($>10^8$ years) outbursts as are the outer radio lobes.  The extended
emission from NGC4478 (labelled in the ROSAT HRI image) and the
southern 37~kpc arc (just outside the outer contour of the southern
radio lobe) are seen in
both the ROSAT-HRI and XMM-Newton images. }
\label{fig:rosat}
\label{fig:mos}
\end{figure*}

\begin{figure}
  \centerline{\includegraphics[width=0.95\linewidth]{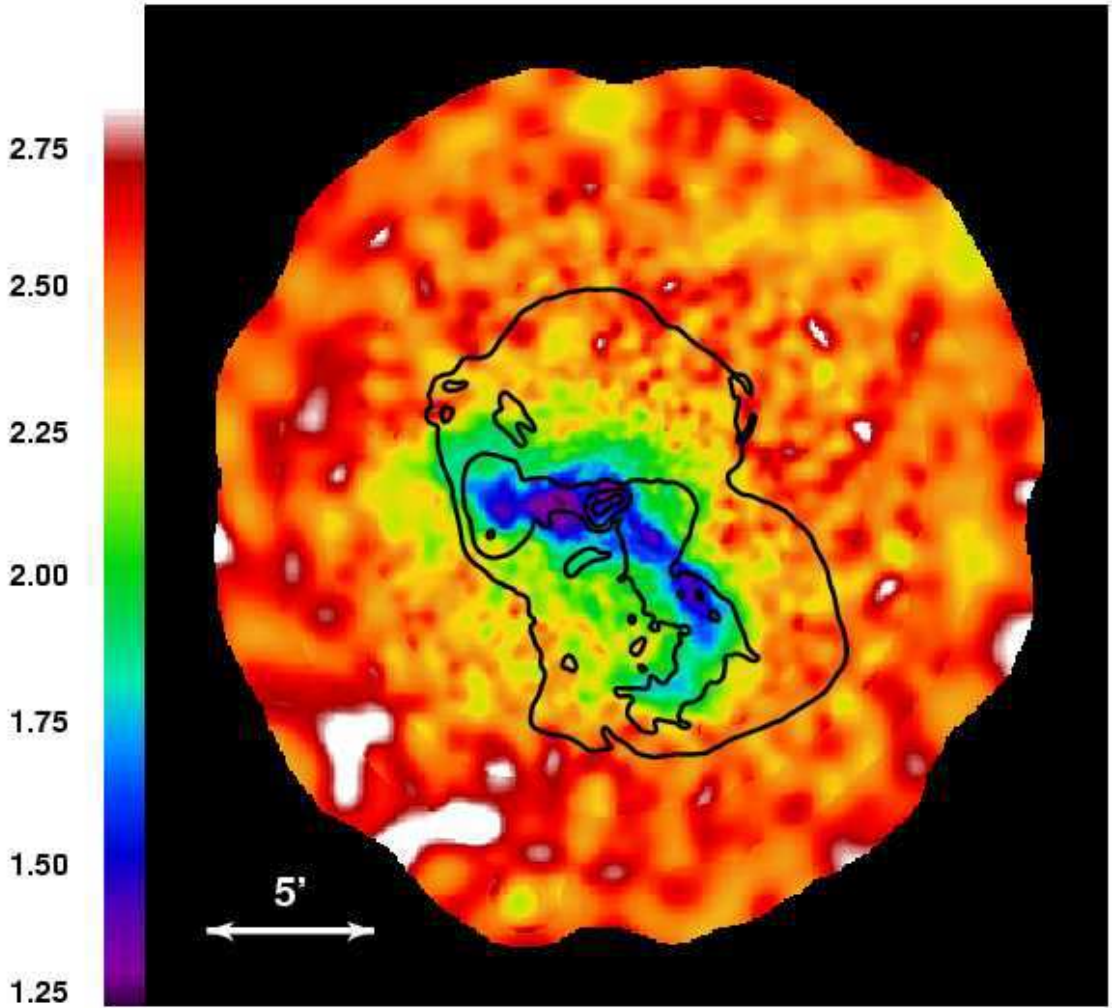}}
  \caption{The XMM-Newton temperature map generated according to the
  method described by Churazov et al. (1996) and summarized in the
  text. The eastern and southwestern arms are cooler than
  the ambient gas (as already discussed in earlier XMM-Newton analyses
  e.g., Belsole et al. 2001 and Molendi 2002). The temperature map is
  adaptively smoothed to reduce the noise and hence small scale
  features are necessarily broadened. Contours from the 90~cm image
  from Owen et al. are superposed (0.25, 2.5, 25, $250\times10^{-3}$
  JY per $1.5''\times1.5''$ pixel)}. 
\label{fig:xmm_tmap}
\end{figure}

\begin{figure*}[t]
  \centerline{
  \includegraphics[width=0.3\linewidth]{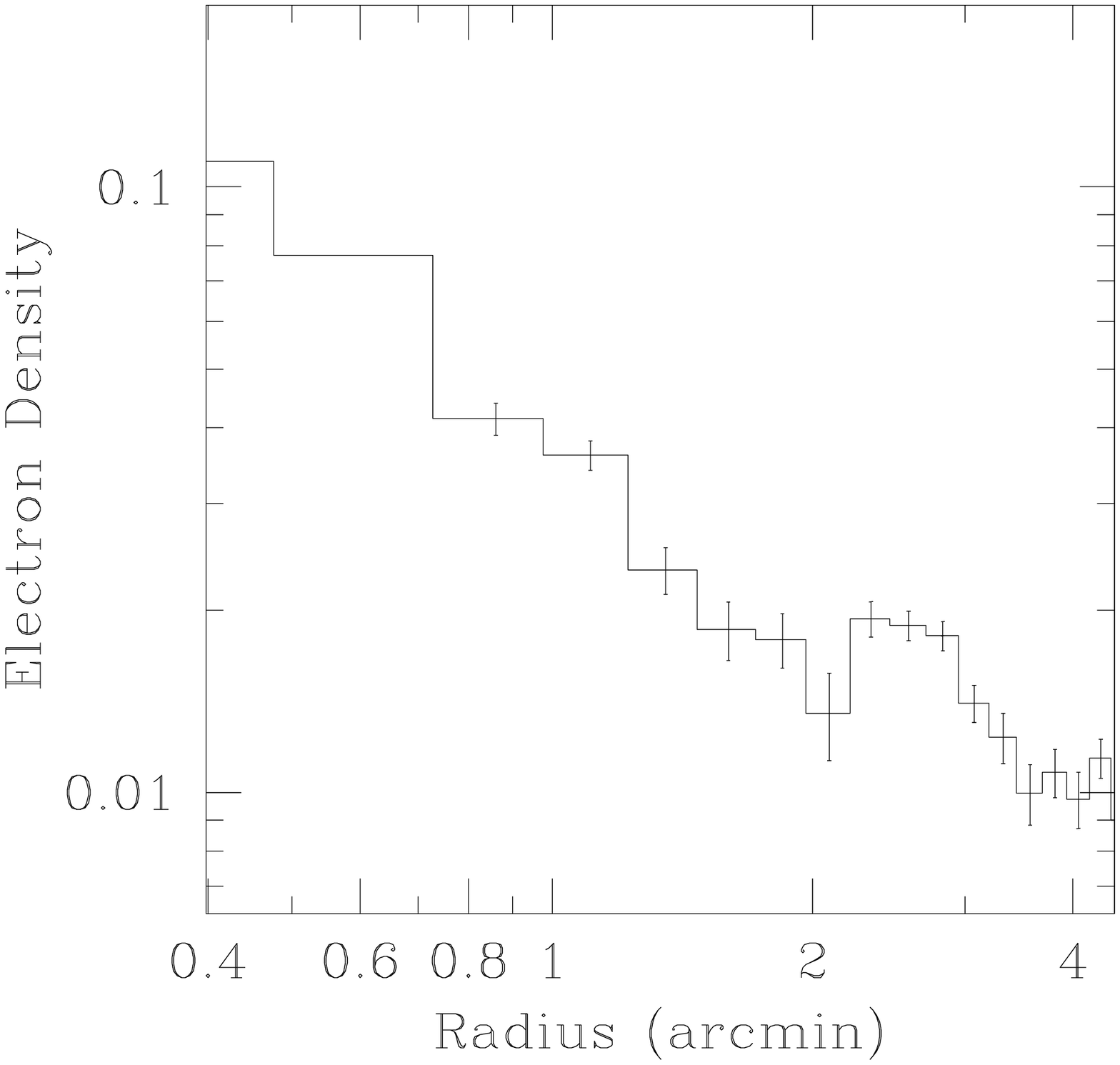}
  \includegraphics[width=0.3\linewidth]{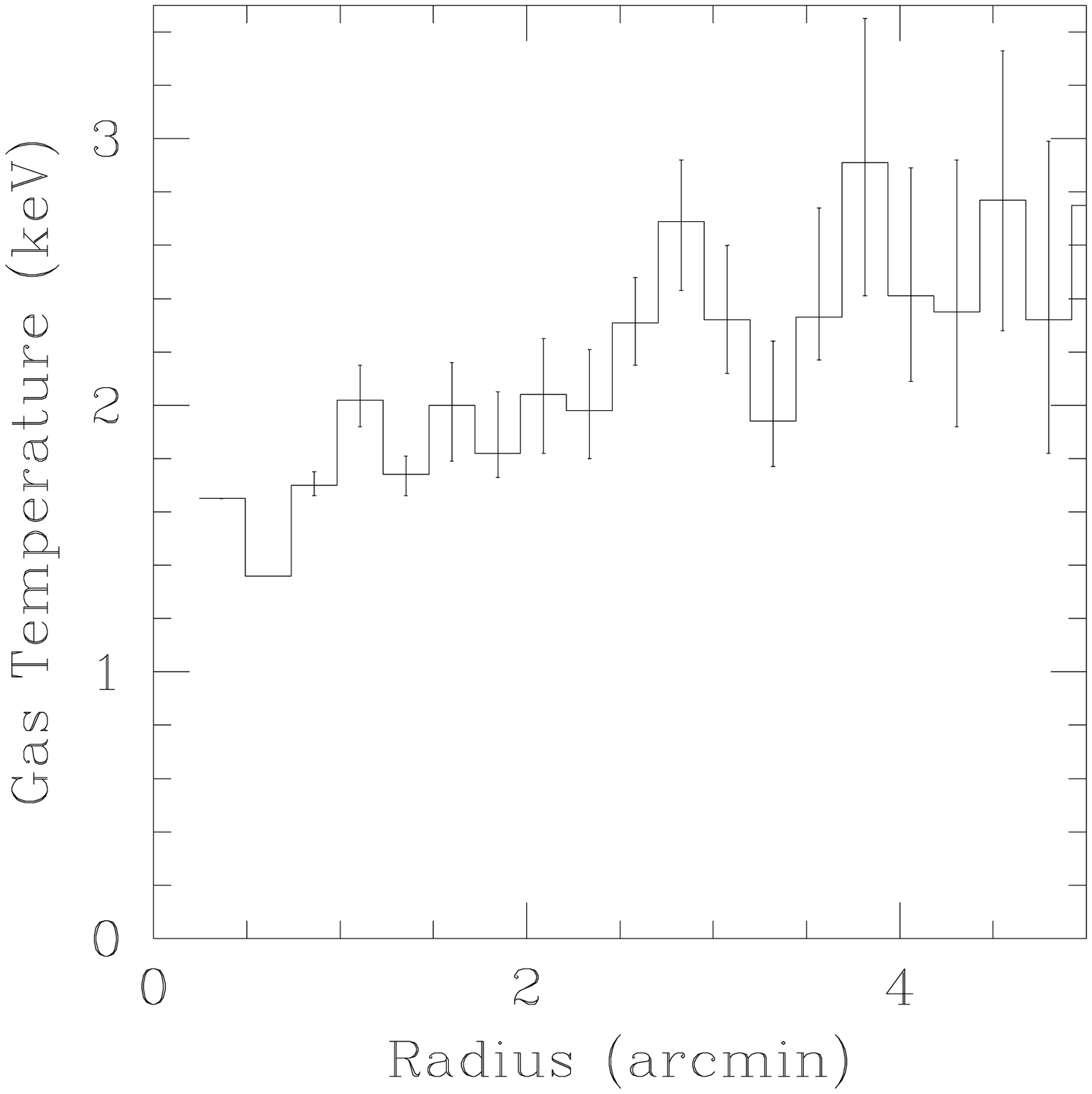}
  \includegraphics[width=0.3\linewidth]{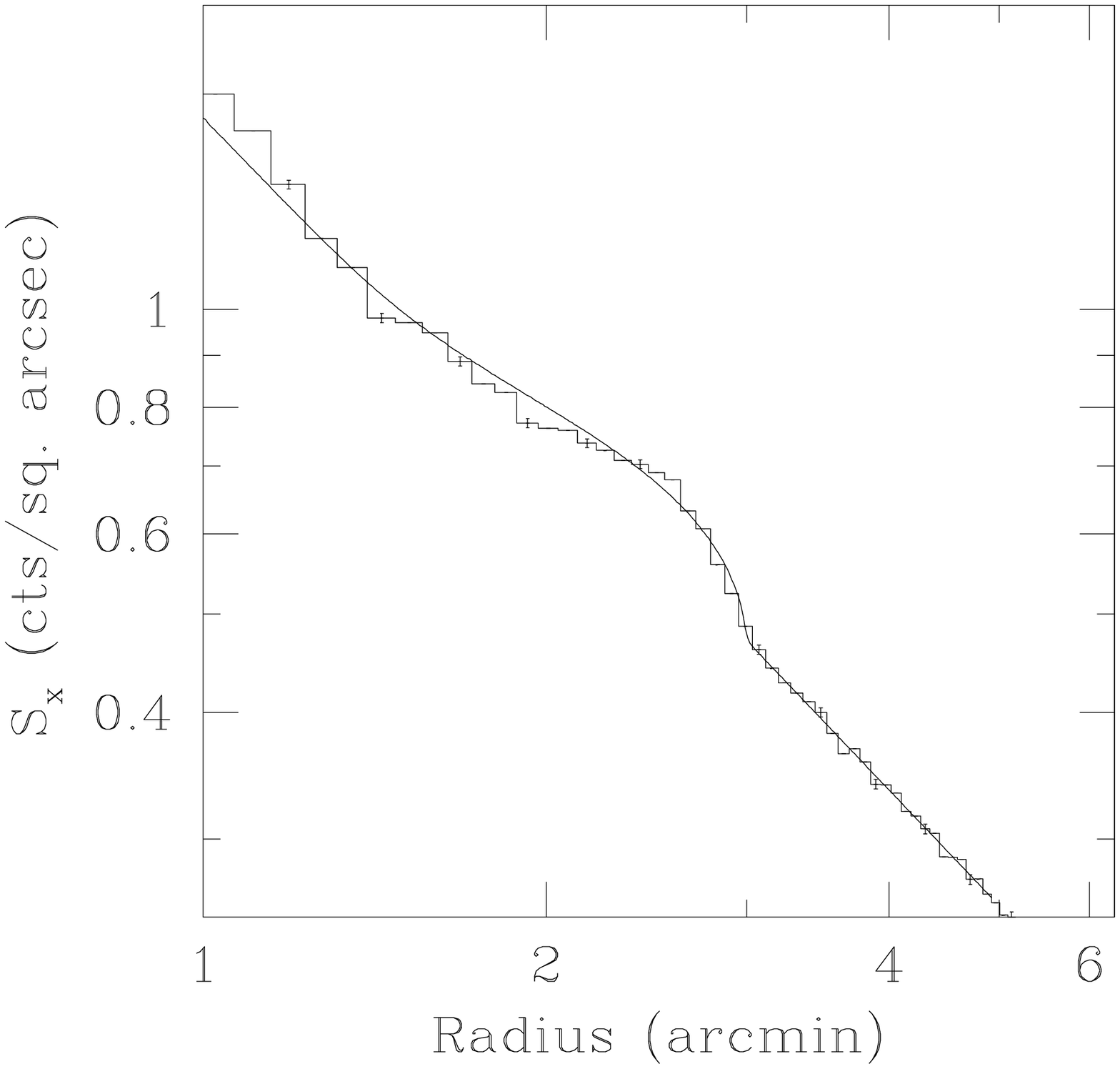}
}
\caption{(Left) Radial electron density profile shows the 14~kpc ($3'$)
  ring. (Center) Deprojected gas temperature.  (Right) Model compared
  to the observed surface brightness profile.  The deprojected gas
  density and temperature are derived by fitting the outermost bin and
  then using the fit results, weighted by the projected emissivity of
  the outermost ring, as one component of the fit for the next inner
  ring. Repeating this process inward yields the deprojected
  temperature and density profile. We use the deprojected values of
  gas density and temperature to calculate gas pressures.
}  \label{fig:ring_model}
\end{figure*}

\begin{figure}
  \centerline{\includegraphics[width=0.85\linewidth]{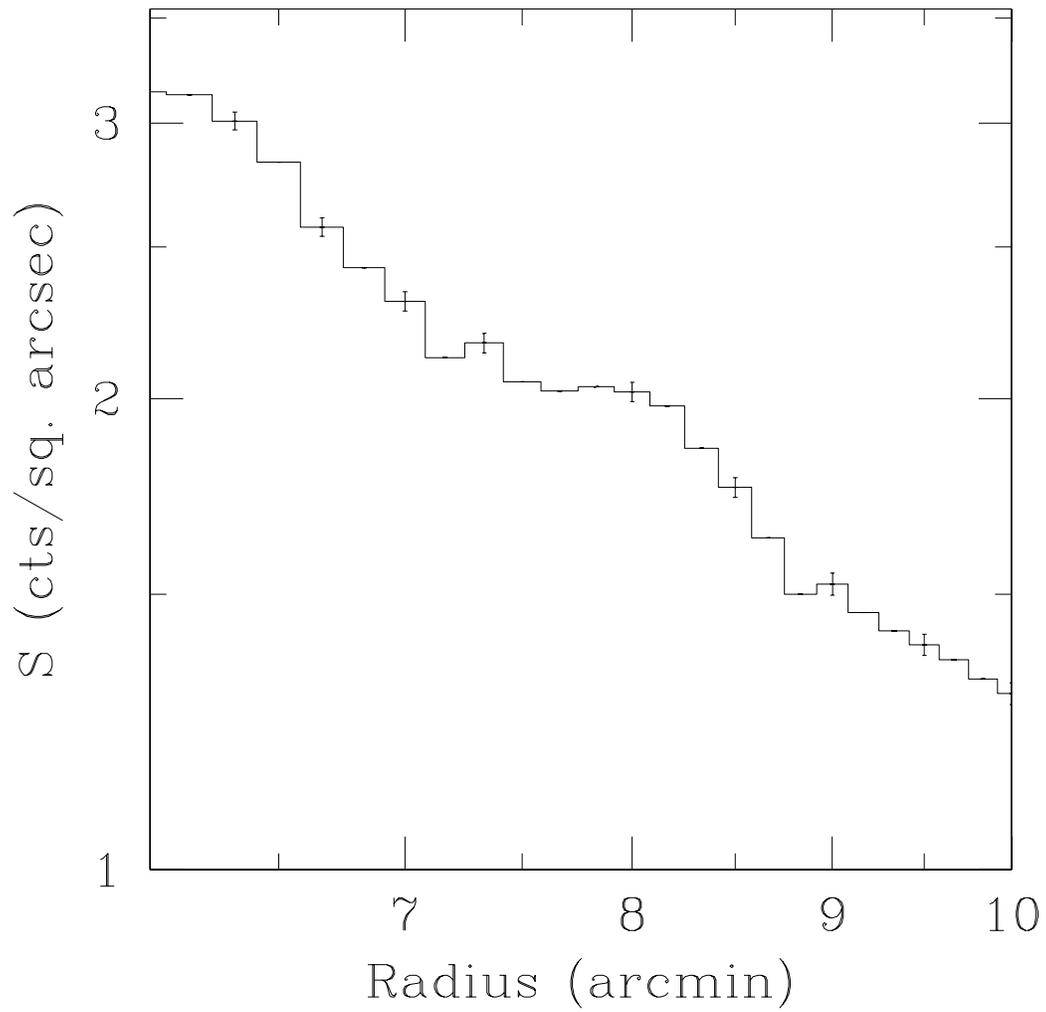}}
  \caption{A radial profile extracted from the merged Chandra image
  (0.5-2.5 keV) 
  shows the excess emission 37~kpc from the nucleus of M87. The profile was
  extracted over an azimuth of 30\deg.}
\label{fig:south_profile}
\end{figure}

\begin{figure}
  \centerline{\includegraphics[width=0.95\linewidth]{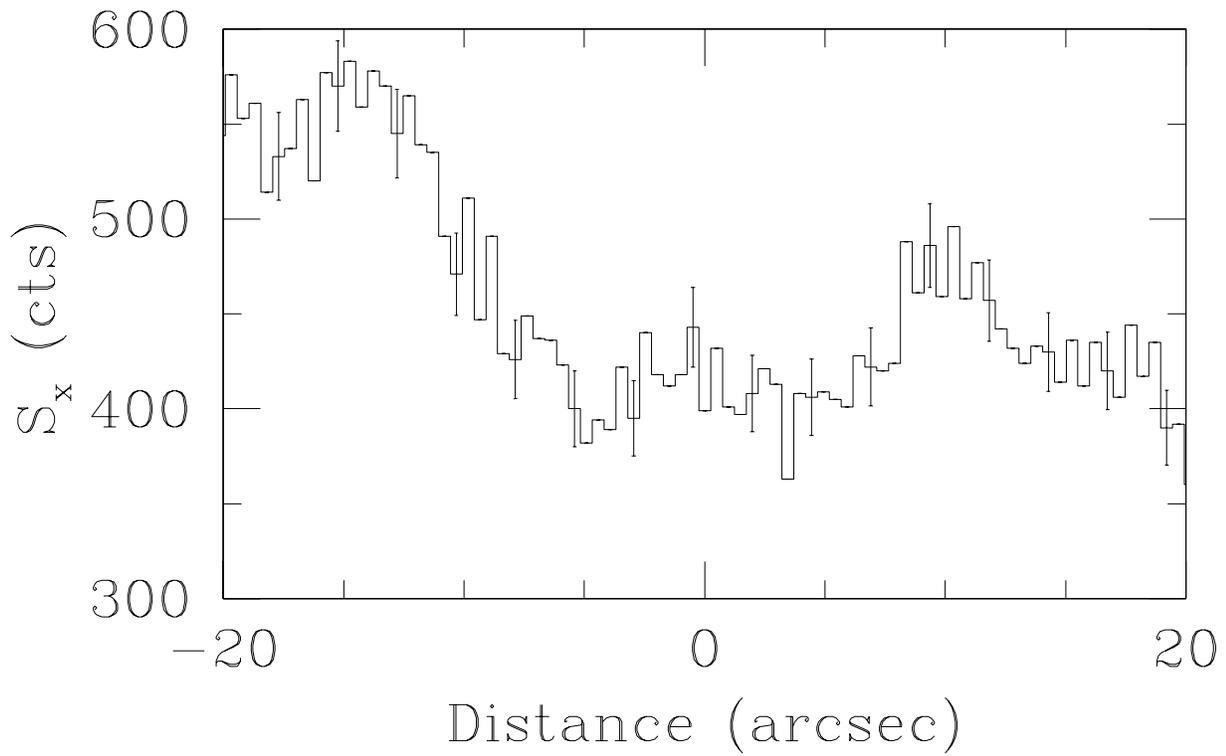}}
  \caption{A projection across one of the ``effervescent'' bubbles at
  the base of the eastern arm. This particular bubble is centered
  $1.25'$ (5.8~kpc) east of the M87 nucleus (and is labeled ``bubble''
  in Fig.~\ref{fig:bl1sum}c). The width of the
  projection is the bubble diameter, $20''$ (1.6~kpc). }
\label{fig:bubble_proj}
\end{figure}

\begin{figure}
  \centerline{\includegraphics[width=0.95\linewidth]{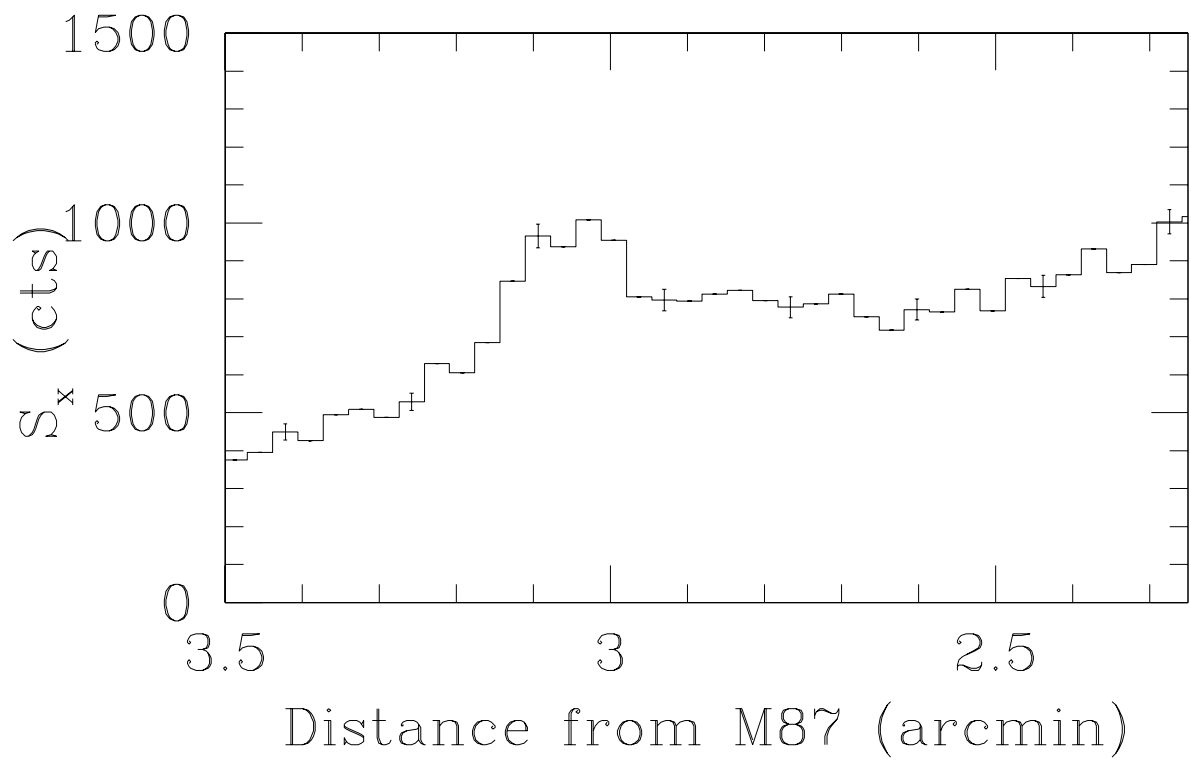}}
  \caption{A projection along the eastern arm shows the sharp surface
  brightness enhancement at a radius of $\sim3'$ comparable to that of
  the 14~kpc ring. A similar brightening is seen at
  approximately the same distance from the M87 nucleus along the
  southwestern arm (see Fig.~\ref{fig:divking}).}
\label{fig:eastern_arm_proj}
\end{figure}

\begin{figure} 
  \centerline{\includegraphics[width=0.80\linewidth]{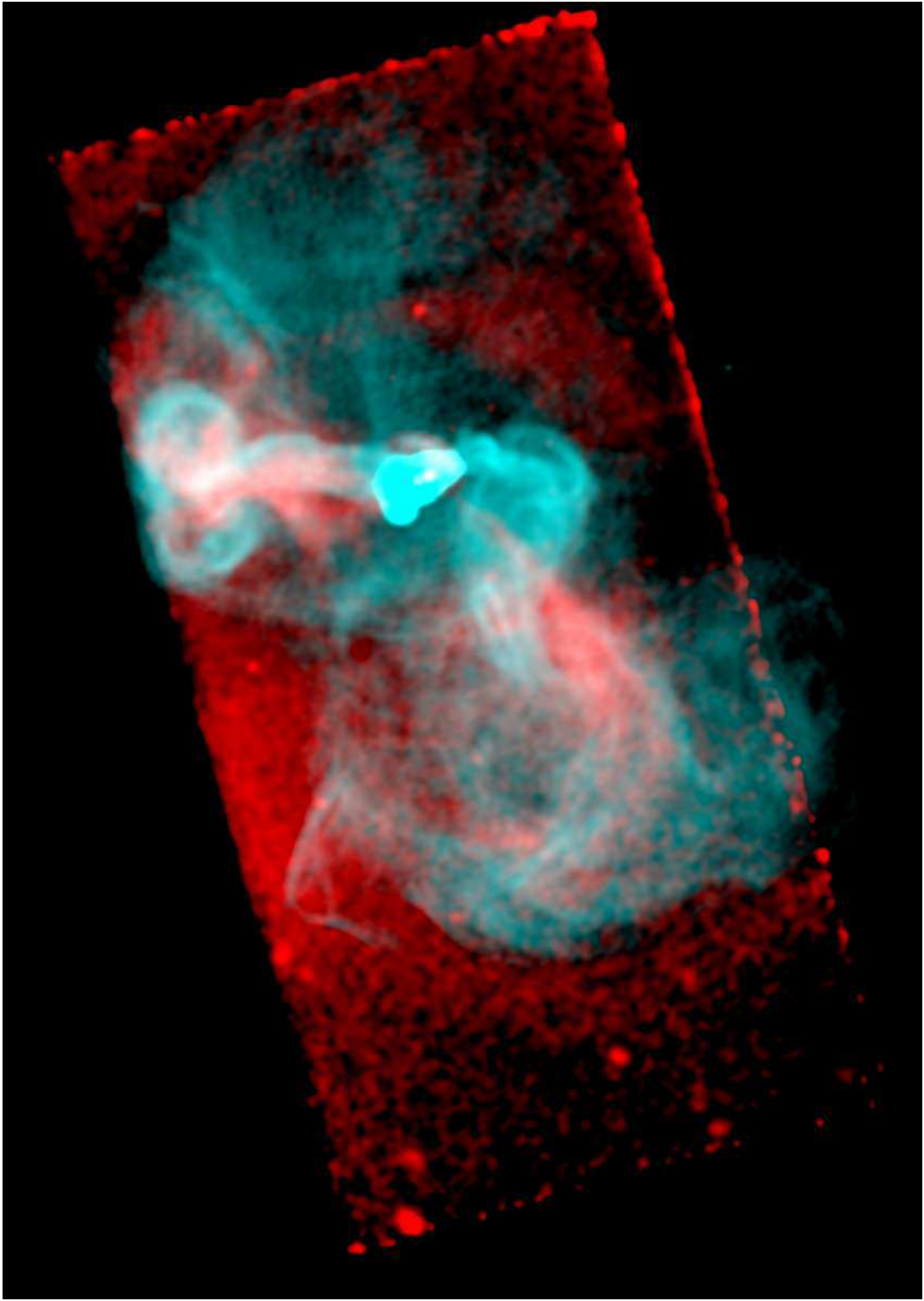}}
  \caption{This figure combines the X-ray image (red) from
  Fig.~\ref{fig:divking} with the 90~cm radio map (light blue) of Owen et
  al. (2000) and emphasizes both the similarities and differences
  between the X-ray and radio emission. In the eastern arm, the X-ray
  and radio appear nearly coincident as one might expect for a cool,
  dense X-ray column generated by buoyant bubbles. The end of the radio
  arm shows the clear toroidal shape expected from a large buoyant
  bubble. To the southwest, the radio emission appears to spiral
  around the narrow X-ray filament until the X-ray filament divides
  and bends to the east.}
\label{fig:overlay}
\end{figure}

\end{document}